\documentclass[aps,prl,reprint,twocolumn,floatfix,hidelinks,superscriptaddress,nofootinbib]{revtex4-2}

\usepackage{braket}  
\usepackage{tikz}
\usepackage{pgfplots}
\pgfplotsset{compat=newest}
\usetikzlibrary{calc, patterns, angles, arrows.meta, bending, shadings, math, shapes}

\usepackage{graphicx, xcolor}
\usepackage{siunitx}
\usepackage{amsmath, amssymb}
\usepackage{csquotes}
\usepackage[hidelinks]{hyperref}
\usepackage[nolist,smaller]{acronym}

\definecolor{mlblue}{rgb}{0,0.4470,0.7410}
\definecolor{mlred}{rgb}{0.8500,0.3250,0.0980}
\definecolor{mlorange}{rgb}{0.9290,0.6940,0.1250}
\definecolor{mlviolet}{rgb}{0.4940,0.1840,0.5560}
\definecolor{mlgreen}{rgb}{0.4660,0.6740,0.1880}
\definecolor{mldarkred}{rgb}{0.6350,0.0780,0.1840}
\definecolor{mlcyan}{rgb}{0.3010,0.7450,0.9330}

\newcommand{\Title}{Experimental Quantum Advantage in the Odd-Cycle Game}

\newcommand{\ion}[2]{\mbox{$^{#2}$#1$^+$}}

\newcommand{\Sr}{\ion{Sr}{88}}
\newcommand{\lev}[2]{{#1}_{#2}}
\newcommand{\fslev}[3]{{#1}_{#2},\,m_{J}\!=\!{#3}}  

\newcommand{\ish}{\mbox{$\sim$}\,}

\begin{document}

\title{\Title{}}
\author{P.~Drmota}
\author{D.~Main}
\author{E.~M.~Ainley}
\author{A.~Agrawal}
\author{G.~Araneda}
\author{D.~P.~Nadlinger}
\author{B.~C.~Nichol}
\author{R.~Srinivas}
\affiliation{Department of Physics, University of Oxford, Clarendon Laboratory, Parks Road, Oxford OX1 3PU, U.K.}

\author{A.~Cabello}
\affiliation{Departamento de F\'isica Aplicada II, Universidad de Sevilla, E-41012 Sevilla, Spain}
\affiliation{Instituto Carlos I de F\'isica Te\'orica y Computacional, Universidad de Sevilla, E-41012 Sevilla, Spain}

\author{D.~M.~Lucas}
\affiliation{Department of Physics, University of Oxford, Clarendon Laboratory, Parks Road, Oxford OX1 3PU, U.K.}
\date{\today}


\begin{abstract}
    We report the first experimental demonstration of the odd-cycle game.
    We entangle two atoms separated by \ish\SI{2}{\meter} and the players use them to win the odd-cycle game with a probability $\sim 26\sigma$ above that allowed by the best classical strategy.
    The experiment implements the optimal quantum strategy, is free of loopholes, and achieves \SI{97.8(3)}{\percent} of the theoretical limit to the quantum winning probability.
    We perform the associated Bell test and measure a nonlocal content of 0.54(2) -- the largest value for physically separate devices, free of the detection loophole, ever observed.
\end{abstract}


\maketitle

%
Quantum advantage refers to the use of quantum systems to perform a given task more efficiently than classical physics allows.
But what does it take to convince {\em anyone} that quantum advantage is real?
We suggest answering this question by identifying a problem (i) that can be expressed in everyday terms, and (ii) whose \textit{optimal} classical solution is evident without a formal mathematical argument, and proposing an experiment (iii) that is \textit{feasible}, (iv) that is a \textit{faithful} implementation of the problem, and (v) whose result will \textit{immediately} convince sceptics.

The quantum advantage for random circuit sampling~\cite{Arute:2019NAT} and Gaussian boson sampling~\cite{Zhong:2021PRL} is impressive, but these problems involve complex probability distributions, and their classical solutions have not yet been proven optimal~\cite{pan_solving_2022, clifford_faster_2024}.
Even the violation of a Bell inequality~\cite{Clauser:1969PRL} may not suffice for our purpose, since the classical limit of the inequality may not be evident without formal mathematical proof.
One candidate could be the three-player Greenberger-Horne-Zeilinger (GHZ) game~\cite{steane_physicists_2000, Mermin:1990AJP}.
However, the state-of-the-art fidelity of GHZ states between three separated systems~\cite{pompili_realization_2021} does not yet permit quantum advantage at this game.
We can give an entangled photon to each player~\cite{Pan:2000NAT}, but then we have to contend with the detection loophole~\cite{Cabello:2008PRL} and add some post-processing.
As we shall see, the odd-cycle game~\cite{cleve_consequences_2004}, however, arguably satisfies both (i) and (ii).
In this Letter we present a realisation of the optimal quantum strategy to win this game.
Our implementation satisfies (iii), (iv), and (v), as the game is played by independent, physically separated players, and is free of loopholes (no extra assumptions are needed).

\begin{figure}[ht]
    \includegraphics{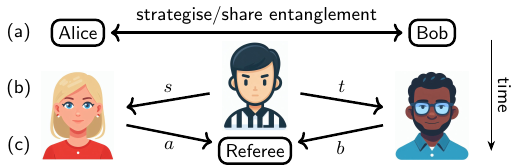}
    \caption{
        One round of a nonlocal game consists of three steps.
        (a) The players strategise and/or share entanglement.
        (b) A referee provides inputs ($s, t$) to the players.
        (c) The players independently output $a$ and $b$, and the referee evaluates the winning condition for that round.
        Many rounds are repeated to estimate the winning probability for each strategy.
    }\label{fig:nonlocal game}
\end{figure}

Bipartite nonlocal games~\cite{brassard_quantum_2005} involve two cooperating players, Alice and Bob, interacting with a referee (Fig.~\ref{fig:nonlocal game}).
The players can choose a strategy involving quantum entanglement to increase their winning chances compared to a local, classical strategy.
Nonlocal games differ from bipartite Bell tests in the way the local measurement settings are chosen.
In bipartite Bell tests, the parties are assumed to have free will and choose their inputs independently.
In nonlocal games, the rules include that a referee chooses all inputs (queries) for all players according to a predefined joint distribution and that the players return their outputs without consulting each other.
The referee then evaluates the player's responses to the inputs using a winning condition.

This definition of nonlocal games precludes some of the loopholes that typically bedevil bipartite Bell tests.
For instance, the freedom-of-choice assumption becomes obsolete because the referee chooses the settings instead of the players.
Moreover, as no communication is allowed between players and all inputs originate from the referee, it is neither necessary nor possible to impose spacelike separation between the input of one player and the output of another player, unlike in loophole-free Bell tests~\cite{hensen_loophole-free_2015, shalm_strong_2015, giustina_significant_2015, rosenfeld_event-ready_2017, storz_loophole-free_2023}.

A nonlocal game exhibits quantum advantage if, when players share entanglement, they achieve a winning probability $\omega_\mathrm{q}$, that is greater than the probability $\omega_\mathrm{c}$ of winning using the best classical strategy.
For the simplest nonlocal  game, the Clauser-Horne-Shimony-Holt (CHSH) game~\cite{Clauser:1969PRL}, the quantum strategy wins on average $\omega_\mathrm{q}=\cos^2(\pi/8)\approx\SI{85}{\percent}$ of rounds compared to the best classical strategy with $\omega_\mathrm{c} = \SI{75}{\percent}$.
Without knowledge of quantum physics, an observer would be drawn to conclude that the players communicate telepathically in order to win the game more frequently than mathematically possible.
For this reason, \enquote{pseudo-telepathy} is attributed to games for which $\omega_\mathrm{q} = 1$, such as the magic square game~\cite{Cabello:2001PRLa,Cabello:2001PRLb,Aravind:2004AJP,brassard_quantum_2005}.
However, it can be proven that the quantum advantage of any bipartite nonlocal game with an output alphabet of size 2, such as the CHSH game and the odd-cycle game, is restricted to $\omega_\mathrm{q} < 1$~\cite{brassard_minimum_2004}.

\begin{figure}[t]
    \centering
    \includegraphics{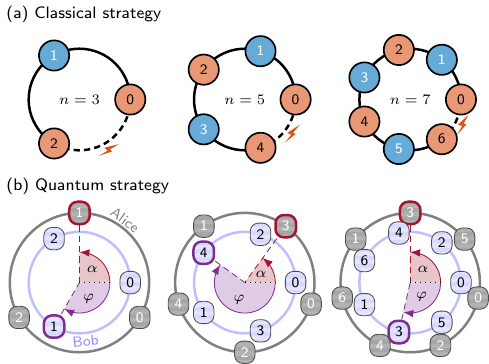}
    \caption{
        Two cooperating players, Alice and Bob, try to convince a referee that the vertices of an odd-cycle can be two-coloured.
        The best classical strategy (a) is for the players to agree on one particular colouring before the game (examples shown for $n\in\{3,5,7\}$).
        This strategy wins all queries except the one where adjacent vertices inevitably have the same colour (see dashed edges).
        In the quantum strategy (b), Alice and Bob each measure their half of an entangled Bell state.
        The optimal qubit rotation angles [Eq.~\eqref{eq:povms}] for any given query (vertex to specify the colour of) are shown, with example inputs highlighted.
        Alice's angles (outer circle) are shifted by an angle $\pi/(2n)$ with respect to Bob's (inner circle).
        Using the correlated nature of the measurement outcomes, the players can win the game more often than using the optimal classical strategy.
    }\label{fig:strategies}
\end{figure}

Nonlocal games have so far only been demonstrated in a photonic system~\cite{xu_experimental_2022}; earlier photonic experiments explored nonlocality and related concepts~\cite{weihs_violation_1998, zhao_experimental_2003, huang_experimental_2003, pomarico_various_2011, Aolita:2012PRA, Christensen:2015PRX, hu_experimental_2016, qu_state-independent_2021}. 
In those cases, imperfect detection efficiencies made it necessary to ignore no-click outcomes while invoking the fair sampling assumption to acknowledge the possibility of hidden variables biasing the distribution of photon loss events.
This constitutes the detection loophole, which introduces margin for the players to cheat in nonlocal games by claiming photon loss depending on the inputs.

The detection efficiency requirements needed to avoid this loophole are challenging to overcome in photonic systems~\cite{shalm_strong_2015}.
However, atomic systems can be measured with near-perfect efficiency, and have been used to perform detection-loophole-free tests of contextuality, and to violate Bell inequalities~\cite{rowe_experimental_2001, matsukevich_bell_2008, kirchmair_state-independent_2009, rosenfeld_event-ready_2017, tan_chained_2017}.
%
%
Here, we implement the odd-cycle game for the first time and show experimentally that the quantum strategy outperforms the best classical strategy.
We use remotely entangled trapped atomic ions, which are among the leading platforms for quantum networking in terms of entanglement fidelity and rate~\cite{stephenson_high-rate_2020}.
This allows us to argue convincingly that the players are physically separated by a macroscopic distance and that each of them controls and acts on a different physical system in a totally independent way.
Furthermore, as trapped ions can be measured with high efficiency, we do not invoke the fair sampling assumption to justify the quantum advantage.

To illustrate the problem underlying the odd-cycle game, imagine setting a round table with an odd number of seats, $n$, and alternating red and blue plates.
Because there is an odd number of seats, but only 2 colours, there is always one offending adjacent pair of plates with the same colour [Fig~\ref{fig:strategies}(a)].
In the odd-cycle game, a referee individually asks two isolated players, Alice and Bob, to specify the colour for a given seat, providing that they are given either the same seat, or adjacent seats, with Alice's seat, $s$, always to the \enquote{left} of Bob's seat, $t$, such that $t \equiv s + 1\ (\mathrm{mod}\ n)$.
For example, if $n=3$, the set of inputs is $\{00, 01, 11, 12, 22, 20\}$.
In each round, the referee selects one of the $2n$ possible queries with uniform probability.
Each of Alice and Bob return a bit indicating the colour of the plate, back to the referee.
The players win a round of the game if and only if their responses ($a,b$) satisfy the following winning condition: when given the same seat, their response should be the same, and when given adjacent seats, their responses should be different.
Since they are not allowed to communicate during the game, their best classical strategy is to agree on a particular colouring before the game, such as $a = s\ \mathrm{mod}\ n$ and $b = t\ \mathrm{mod}\ n$, as illustrated in Fig.~\ref{fig:strategies}(a) for $n\in\{3,5,7\}$.
The players win all queries except the one where, in their chosen arrangement, adjacent seats inevitably have been assigned the same colour.
Therefore, using the optimal classical strategy, the winning probability is $\omega_\mathrm{c} = 1 - 1/(2n)$.

To implement the optimal quantum strategy~\cite{cleve_consequences_2004} (see Supplemental Material for an alternative proof of optimality), Alice and Bob share a Bell state, {$\ket{\Psi^{+}} = (\ket{0}_\mathrm{A}\ket{1}_\mathrm{B} + \ket{1}_\mathrm{A}\ket{0}_\mathrm{B})/\sqrt{2}$}, prior to each round of the game.
The states $\ket{0}$ and $\ket{1}$ are eigenstates of the Pauli operator $\hat{Z}$.
In the following, we will refer to vertices in an odd cycle instead of seats on a round table.
Once the players have generated entanglement, the referee distributes the queried vertices, $s$ and $t$, to Alice and Bob, who perform a qubit rotation, $\hat{R}_y(\theta) = \cos(\theta/2) \hat{I} - \mathrm{i} \sin(\theta/2) \hat{Y}$ with rotation angles given by
\begin{subequations} \label{eq:povms}
    \begin{align}
        \alpha_s &= \varphi_s - \frac{\pi}{2n} \ , \\
        \beta_t &= -\varphi_t \ ,
    \end{align}
\end{subequations}
respectively, where $\varphi_v = \pi v (n-1)/n$ [Fig.~\ref{fig:strategies}(b)].
Finally, they obtain their responses, $a$ and $b$, by measuring their qubits in the $\hat{Z}$ basis.
Alice inverts her measurement result to obtain her output bit $a$, and Bob uses his measurement result directly as output bit $b$.
This process implements a qubit-qubit strategy~\cite{braunstein_wringing_1990} with a winning probability of $\omega_\mathrm{q}=\cos^2[\pi/(4n)]$.

\begin{figure}[ht]
    \includegraphics{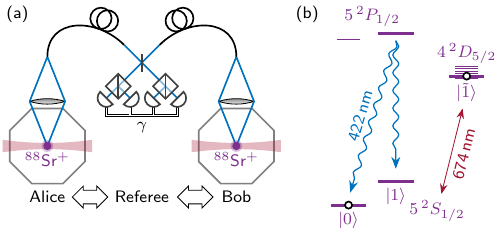}
    \caption{
        (a) Alice and Bob each operate an ion trap loaded with a single \Sr{} ion. The ions are entangled photonically by entanglement swapping at a central heralding station.
        (b) Level diagram of \Sr{} (not to scale). Single photons for generating entanglement are spontaneously emitted on the 422-nm $\lev{S}{1/2}\leftrightarrow\lev{P}{1/2}$ transition. Upon a two-photon herald, the $\lev{S}{1/2}$ ground state qubits $\{\ket{0} = \ket{m_J = -1/2}, \ket{1} = \ket{m_J = 1/2}\}$ of the remote ions become entangled. A 674-nm $\pi$ pulse then maps $\ket{1}$ to $\ket{\tilde{1}} = \ket{\fslev{D}{5/2}{-3/2}}$, and the query-dependent rotations are performed on the resulting optical qubit, $\{\ket{0},\ket{\tilde{1}}\}$.
    }\label{fig:apparatus}
\end{figure}
In our experiment [Fig.~\ref{fig:apparatus}(a)], each player controls their own apparatus for trapping and manipulating a \Sr{} trapped-ion qubit~\cite{stephenson_high-rate_2020}.
The systems are separated by \ish\SI{2}{\meter} and controlled independently.
The referee implements a state machine with synchronised digital communication with the control systems of the two players.
The trapped ions are interfaced to optical fibres using free-space collection optics.
To generate remote entanglement in preparation for a round of the odd-cycle game, both ions are simultaneously excited by a 10-ps laser pulse.
Hereafter, the polarisation of each of the spontaneously emitted 422-nm single photons is entangled with the emitter's spin state [Fig.~\ref{fig:apparatus}(b)].
A photonic Bell state measurement at a central heralding station swaps entanglement from the ion-photon pairs onto the remote ions.
One of four two-photon coincidence events, $\gamma \in \{0,1,2,3\}$, each corresponding to a different photon detector click pattern, heralds the creation of
\begin{align}
    \ket{\Psi^{+}(\vartheta_\gamma)} = \frac{1}{\sqrt{2}}\left(\ket{0}_\mathrm{A}\ket{1}_\mathrm{B} + \mathrm{e}^{\mathrm{i} \vartheta_\gamma} \ket{1}_\mathrm{A}\ket{0}_\mathrm{B}\right) \ .
    \label{eq:psi}
\end{align}
To produce the same state, $\ket{\Psi^{+}}$, independent of $\gamma$, Bob applies the phase gate $\exp(\mathrm{i} \hat{Z} \vartheta_\gamma / 2)$ to his qubit.
Generating entanglement prior to the start of a game takes, on average, \ish\SI{30}{\milli\second};
generation attempts are performed at \SI{1}{\mega\hertz} rate for \SI{250}{\micro\second} with interleaved periods of \SI{770}{\micro\second} consisting of Doppler cooling and electromagnetically-induced transparency cooling of the axial motion.
Compared to~\cite{stephenson_high-rate_2020}, additional ground state cooling and shorter attempt periods increased Bell state fidelities to \ish\num{0.97}.
The subsequent interaction with the referee lasts \ish\SI{600}{\micro\second}, which includes communication of the inputs ($<\SI{1}{\micro\second}$), qubit manipulation (\ish\SI{20}{\micro\second}), qubit readout (\SI{550}{\micro\second}), communication of the outputs ($<\SI{1}{\micro\second}$), and delays (\ish\SI{30}{\micro\second}).  

\begin{figure*}[t]
    \includegraphics{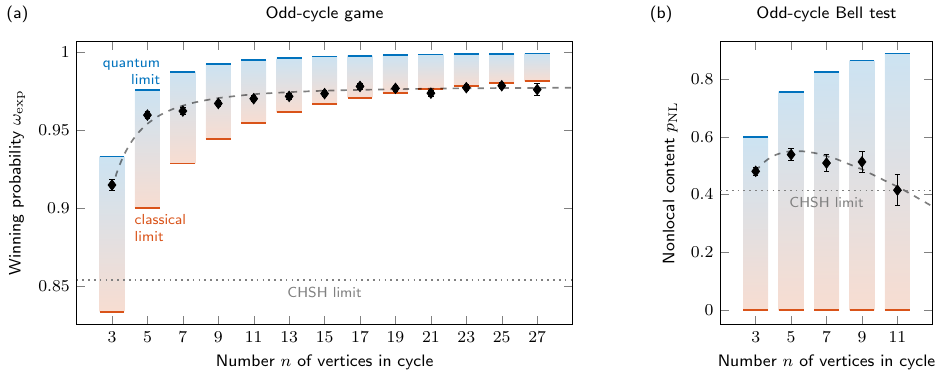}
    \caption{
        Experimental results demonstrating quantum advantage and nonlocality in the odd-cycle game.
        (a) Alice and Bob use a pre-shared entangled state to increase their winning probability compared to the best classical approach. For $n=3$, the winning probability is \ish$26\sigma$ above the classical limit. Due to experimental imperfections, their quantum advantage is limited to $n \leq 19$.
        (b) The nonlocal content $p_\mathrm{NL}$ is measured in a separate Bell experiment, where Alice and Bob choose their inputs independently.
        Error bars represent the binomial standard error.
        Dashed curves show the theoretical prediction for performance at the \SI{97.8}{\percent} level to the quantum limit.
        Dotted lines indicate the quantum limit for the CHSH case for reference.
    }\label{fig:results}
\end{figure*}

The players implement the input-dependent rotation using a sequence of two laser pulses on the 674-nm qubit transition $\ket{0} \leftrightarrow \ket{\tilde{1}}$ [Fig.~\ref{fig:apparatus}(b)] followed by state-dependent fluorescence readout.
First, a $\pi/2$ pulse around the $\hat{X}$ axis effectively swaps the $\hat{Z}$ axis with the $\hat{Y}$ axis.
Then, a $\pi/2$ pulse with a relative phase of $\pi - \theta$ applies the target rotation $\theta \in \{\alpha_s, \beta_t\}$ around the $\hat{Z}$ axis while swapping the $\hat{Z}$ and $\hat{Y}$ axes back.
This way, the variable angle $\theta$ is encoded in a digital phase instead of a physical pulse duration, for which independent calibrations would be necessary.
Both players transmit their outputs electronically in real time to the referee to conclude one round of the game.

We realise the odd-cycle game for $3 \leq n \leq 27$ and measure the experimental winning probability $\omega_\mathrm{exp}$ [Fig.~\ref{fig:results}(a)].
The number of vertices $n$ is announced before the start of the game and the players use it to calculate the measurement angles $\alpha_s$ and $\beta_t$ [Eq.~\eqref{eq:povms}] in every round.
In total, \num{101000} rounds of the odd-cycle game were played.
On average, the players perform at a $\langle\omega_\mathrm{exp}/\omega_\mathrm{q}\rangle=\SI{97.8(3)}{\percent}$ level with respect to the theoretical limit imposed by quantum mechanics.
These values are significantly above the classical limits for $n \leq 19$, thus demonstrating a clear quantum advantage at this game.
The observed reduction compared with the quantum limit is dominated by imperfections in the entangled state [Eq.~\ref{eq:psi}].
The measurement error is independent of the number of vertices $n$ in the odd-cycle and the particular inputs $s$ and $t$ because the qubit rotation angles [Eq.~\eqref{eq:povms}] are encoded in a relative phase.
Contrary to previous experimental demonstrations of nonlocal games using photonic systems~\cite{xu_experimental_2022}, we employ shot-by-shot random switching of the inputs to the players.
Moreover, as the players each have a quantum memory, they can operate in an \enquote{event-ready} manner. Thus, the first stage of the game, in which the players are allowed to \enquote{strategise} [Fig.~\ref{fig:nonlocal game}(a)], is well-separated temporally from the subsequent individual information exchange with the referee [Fig.~\ref{fig:nonlocal game}(b-c)].

In a second experiment, we perform the bipartite Bell test variant of the odd-cycle game by letting the players choose their inputs independently.
We use the Bell inequality associated with the winning probability of the odd-cycle game,
\begin{multline} \label{winprob}
    \omega = \frac{1}{2 n} \sum_{j=0}^{n-1} P(0,0|j,j)+P(1,1|j,j)\\
    +P(0,1|j,j+1)+P(1,0|j,j+1) \le \omega_\mathrm{c} \ .
\end{multline}
The correlations measured by a Bell inequality can be captured by a mixture of local and nonlocal correlations, which are constrained by local hidden variables and nonsignaling, respectively, where only the latter contribute to the violation of the inequality~\cite{Elitzur:1992PLA}.
The local content of \eqref{winprob} is upper bounded by $p_\mathrm{L} \le (1 - \omega)/(1 - \omega_\mathrm{c})$ (see Supplementary Material).
In total, \num{134000} repetitions of the odd-cycle Bell test were performed.
The experimentally measured nonlocal content, $p_\mathrm{NL} = 1-p_\mathrm{L}$ [Fig.~\ref{fig:results}(b)] of up to $0.54(2)$ for $n=5$ is, to the best of our knowledge, the highest detection-loophole-free value ever observed between physically separate devices (see~\cite{Christensen:2015PRX} for experimental values obtained with photons without closing the detection loophole, and~\cite{tan_chained_2017} for the highest nonlocal content with the detection loophole closed, but using a single device).
This value is \ish$6\sigma$ above the theoretical value for the maximum quantum violation of the CHSH inequality, $p_\mathrm{NL} = \sqrt{2} - 1 \approx 0.414$.

The results from both the odd-cycle game and the associated Bell test were obtained using the outcomes from every round commencing with entanglement generation, so that the fair-sampling assumption is not required and the detection loophole fully closed.
This eliminates any extra assumptions from the implementation of the odd-cycle game, rendering it loophole-free.
For a loophole-free measurement of nonlocal content above the CHSH limit, the locality loophole is yet to be closed by separating Alice and Bob further to establish space-like separation between their measurements~\cite{hensen_loophole-free_2015, shalm_strong_2015, giustina_significant_2015, rosenfeld_event-ready_2017, storz_loophole-free_2023}.

The same resources that lead to quantum advantage in the odd-cycle game can be applied to practical scenarios in which collaborating agents cannot communicate, such as the rendezvous task~\cite{brukner_entanglement-assisted_2006, mironowicz_entangled_2023, viola_quantum_2024, tucker_quantum-assisted_2024}.
More complex nonlocal games, such as the Kochen-Specker game and the magic square game~\cite{cleve_consequences_2004, xu_experimental_2022}, could be performed in our apparatus if the dimension of the entangled state is increased; by co-trapping a second ion~\cite{drmota_robust_2023}, the players could distribute two Bell pairs at the start of each round, and perform local operations to generate qutrit-qutrit entanglement, or four-qubit entanglement, to implement different quantum strategies.
While the necessary experimental tools for these processes are available in our system~\cite{mixed_species_entanglement}, their fidelity would still need to be improved to demonstrate a quantum advantage at those games.

In summary, we have implemented the odd-cycle game, which is a nonlocal game where a qubit-qubit strategy offers quantum advantage compared to the best classical strategy.
Using remotely entangled trapped ions, we have realised a quantum advantage for odd-cycles of up to $n=19$ vertices.
Since each round produces a valid set of outcomes, we are able to close the detection loophole in this demonstration.
In addition, we have performed the Bell test associated with the odd-cycle game and measured the highest nonlocal content for physically separate systems where the detection loophole is closed.

We thank Sandia National Laboratories for supplying the ion traps used in this experiment, and the developers of the control system ARTIQ~\cite{ARTIQ}.
DPN acknowledges support from Merton College, Oxford.
DM acknowledges support from the U.S.\ Army Research Office (ref.\ W911NF-18-1-0340).
EMA acknowledges support from the U.K.\ EPSRC \enquote{Quantum Communications} Hub EP/T001011/1.
GA acknowledges support from Wolfson College, Oxford.
RS is partially employed by Oxford Ionics Ltd and acknowledges funding from an EPSRC Fellowship EP/W028026/1 and Balliol College, Oxford.
AC is supported by the E.U.-funded project \href{10.3030/101070558}{FoQaCiA} \enquote{Foundations of Quantum Computational Advantage} and the \href{10.13039/501100011033}{MCINN/AEI} (Project No.\ PID2020-113738GB-I00).
This work was supported by the U.K.\ EPSRC \enquote{Quantum Computing and Simulation} Hub EP/T001062/1.

\bibliography{library}

\begin{thebibliography}{47}%
\makeatletter
\providecommand \@ifxundefined [1]{%
 \@ifx{#1\undefined}
}%
\providecommand \@ifnum [1]{%
 \ifnum #1\expandafter \@firstoftwo
 \else \expandafter \@secondoftwo
 \fi
}%
\providecommand \@ifx [1]{%
 \ifx #1\expandafter \@firstoftwo
 \else \expandafter \@secondoftwo
 \fi
}%
\providecommand \natexlab [1]{#1}%
\providecommand \enquote  [1]{``#1''}%
\providecommand \bibnamefont  [1]{#1}%
\providecommand \bibfnamefont [1]{#1}%
\providecommand \citenamefont [1]{#1}%
\providecommand \href@noop [0]{\@secondoftwo}%
\providecommand \href [0]{\begingroup \@sanitize@url \@href}%
\providecommand \@href[1]{\@@startlink{#1}\@@href}%
\providecommand \@@href[1]{\endgroup#1\@@endlink}%
\providecommand \@sanitize@url [0]{\catcode `\\12\catcode `\$12\catcode
  `\&12\catcode `\#12\catcode `\^12\catcode `\_12\catcode `\%12\relax}%
\providecommand \@@startlink[1]{}%
\providecommand \@@endlink[0]{}%
\providecommand \url  [0]{\begingroup\@sanitize@url \@url }%
\providecommand \@url [1]{\endgroup\@href {#1}{\urlprefix }}%
\providecommand \urlprefix  [0]{URL }%
\providecommand \Eprint [0]{\href }%
\providecommand \doibase [0]{https://doi.org/}%
\providecommand \selectlanguage [0]{\@gobble}%
\providecommand \bibinfo  [0]{\@secondoftwo}%
\providecommand \bibfield  [0]{\@secondoftwo}%
\providecommand \translation [1]{[#1]}%
\providecommand \BibitemOpen [0]{}%
\providecommand \bibitemStop [0]{}%
\providecommand \bibitemNoStop [0]{.\EOS\space}%
\providecommand \EOS [0]{\spacefactor3000\relax}%
\providecommand \BibitemShut  [1]{\csname bibitem#1\endcsname}%
\let\auto@bib@innerbib\@empty
\bibitem [{\citenamefont {Arute}\ \emph {et~al.}(2019)\citenamefont {Arute},
  \citenamefont {Arya}, \citenamefont {Babbush}, \citenamefont {Bacon},
  \citenamefont {Bardin}, \citenamefont {Barends}, \citenamefont {Biswas},
  \citenamefont {Boixo}, \citenamefont {Brandao}, \citenamefont {Buell},
  \citenamefont {Burkett}, \citenamefont {Chen}, \citenamefont {Chen},
  \citenamefont {Chiaro}, \citenamefont {Collins}, \citenamefont {Courtney},
  \citenamefont {Dunsworth}, \citenamefont {Farhi}, \citenamefont {Foxen},
  \citenamefont {Fowler}, \citenamefont {Gidney}, \citenamefont {Giustina},
  \citenamefont {Graff}, \citenamefont {Guerin}, \citenamefont {Habegger},
  \citenamefont {Harrigan}, \citenamefont {Hartmann}, \citenamefont {Ho},
  \citenamefont {Hoffmann}, \citenamefont {Huang}, \citenamefont {Humble},
  \citenamefont {Isakov}, \citenamefont {Jeffrey}, \citenamefont {Jiang},
  \citenamefont {Kafri}, \citenamefont {Kechedzhi}, \citenamefont {Kelly},
  \citenamefont {Klimov}, \citenamefont {Knysh}, \citenamefont {Korotkov},
  \citenamefont {Kostritsa}, \citenamefont {Landhuis}, \citenamefont
  {Lindmark}, \citenamefont {Lucero}, \citenamefont {Lyakh}, \citenamefont
  {Mandr{\`a}}, \citenamefont {McClean}, \citenamefont {McEwen}, \citenamefont
  {Megrant}, \citenamefont {Mi}, \citenamefont {Michielsen}, \citenamefont
  {Mohseni}, \citenamefont {Mutus}, \citenamefont {Naaman}, \citenamefont
  {Neeley}, \citenamefont {Neill}, \citenamefont {Niu}, \citenamefont {Ostby},
  \citenamefont {Petukhov}, \citenamefont {Platt}, \citenamefont {Quintana},
  \citenamefont {Rieffel}, \citenamefont {Roushan}, \citenamefont {Rubin},
  \citenamefont {Sank}, \citenamefont {Satzinger}, \citenamefont {Smelyanskiy},
  \citenamefont {Sung}, \citenamefont {Trevithick}, \citenamefont
  {Vainsencher}, \citenamefont {Villalonga}, \citenamefont {White},
  \citenamefont {Yao}, \citenamefont {Yeh}, \citenamefont {Zalcman},
  \citenamefont {Neven},\ and\ \citenamefont {Martinis}}]{Arute:2019NAT}%
  \BibitemOpen
  \bibfield  {author} {\bibinfo {author} {\bibfnamefont {F.}~\bibnamefont
  {Arute}}, \bibinfo {author} {\bibfnamefont {K.}~\bibnamefont {Arya}},
  \bibinfo {author} {\bibfnamefont {R.}~\bibnamefont {Babbush}}, \bibinfo
  {author} {\bibfnamefont {D.}~\bibnamefont {Bacon}}, \bibinfo {author}
  {\bibfnamefont {J.~C.}\ \bibnamefont {Bardin}}, \bibinfo {author}
  {\bibfnamefont {R.}~\bibnamefont {Barends}}, \bibinfo {author} {\bibfnamefont
  {R.}~\bibnamefont {Biswas}}, \bibinfo {author} {\bibfnamefont
  {S.}~\bibnamefont {Boixo}}, \bibinfo {author} {\bibfnamefont {F.~G. S.~L.}\
  \bibnamefont {Brandao}}, \bibinfo {author} {\bibfnamefont {D.~A.}\
  \bibnamefont {Buell}}, \bibinfo {author} {\bibfnamefont {B.}~\bibnamefont
  {Burkett}}, \bibinfo {author} {\bibfnamefont {Y.}~\bibnamefont {Chen}},
  \bibinfo {author} {\bibfnamefont {Z.}~\bibnamefont {Chen}}, \bibinfo {author}
  {\bibfnamefont {B.}~\bibnamefont {Chiaro}}, \bibinfo {author} {\bibfnamefont
  {R.}~\bibnamefont {Collins}}, \bibinfo {author} {\bibfnamefont
  {W.}~\bibnamefont {Courtney}}, \bibinfo {author} {\bibfnamefont
  {A.}~\bibnamefont {Dunsworth}}, \bibinfo {author} {\bibfnamefont
  {E.}~\bibnamefont {Farhi}}, \bibinfo {author} {\bibfnamefont
  {B.}~\bibnamefont {Foxen}}, \bibinfo {author} {\bibfnamefont
  {A.}~\bibnamefont {Fowler}}, \bibinfo {author} {\bibfnamefont
  {C.}~\bibnamefont {Gidney}}, \bibinfo {author} {\bibfnamefont
  {M.}~\bibnamefont {Giustina}}, \bibinfo {author} {\bibfnamefont
  {R.}~\bibnamefont {Graff}}, \bibinfo {author} {\bibfnamefont
  {K.}~\bibnamefont {Guerin}}, \bibinfo {author} {\bibfnamefont
  {S.}~\bibnamefont {Habegger}}, \bibinfo {author} {\bibfnamefont {M.~P.}\
  \bibnamefont {Harrigan}}, \bibinfo {author} {\bibfnamefont {M.~J.}\
  \bibnamefont {Hartmann}}, \bibinfo {author} {\bibfnamefont {A.}~\bibnamefont
  {Ho}}, \bibinfo {author} {\bibfnamefont {M.}~\bibnamefont {Hoffmann}},
  \bibinfo {author} {\bibfnamefont {T.}~\bibnamefont {Huang}}, \bibinfo
  {author} {\bibfnamefont {T.~S.}\ \bibnamefont {Humble}}, \bibinfo {author}
  {\bibfnamefont {S.~V.}\ \bibnamefont {Isakov}}, \bibinfo {author}
  {\bibfnamefont {E.}~\bibnamefont {Jeffrey}}, \bibinfo {author} {\bibfnamefont
  {Z.}~\bibnamefont {Jiang}}, \bibinfo {author} {\bibfnamefont
  {D.}~\bibnamefont {Kafri}}, \bibinfo {author} {\bibfnamefont
  {K.}~\bibnamefont {Kechedzhi}}, \bibinfo {author} {\bibfnamefont
  {J.}~\bibnamefont {Kelly}}, \bibinfo {author} {\bibfnamefont {P.~V.}\
  \bibnamefont {Klimov}}, \bibinfo {author} {\bibfnamefont {S.}~\bibnamefont
  {Knysh}}, \bibinfo {author} {\bibfnamefont {A.}~\bibnamefont {Korotkov}},
  \bibinfo {author} {\bibfnamefont {F.}~\bibnamefont {Kostritsa}}, \bibinfo
  {author} {\bibfnamefont {D.}~\bibnamefont {Landhuis}}, \bibinfo {author}
  {\bibfnamefont {M.}~\bibnamefont {Lindmark}}, \bibinfo {author}
  {\bibfnamefont {E.}~\bibnamefont {Lucero}}, \bibinfo {author} {\bibfnamefont
  {D.}~\bibnamefont {Lyakh}}, \bibinfo {author} {\bibfnamefont
  {S.}~\bibnamefont {Mandr{\`a}}}, \bibinfo {author} {\bibfnamefont {J.~R.}\
  \bibnamefont {McClean}}, \bibinfo {author} {\bibfnamefont {M.}~\bibnamefont
  {McEwen}}, \bibinfo {author} {\bibfnamefont {A.}~\bibnamefont {Megrant}},
  \bibinfo {author} {\bibfnamefont {X.}~\bibnamefont {Mi}}, \bibinfo {author}
  {\bibfnamefont {K.}~\bibnamefont {Michielsen}}, \bibinfo {author}
  {\bibfnamefont {M.}~\bibnamefont {Mohseni}}, \bibinfo {author} {\bibfnamefont
  {J.}~\bibnamefont {Mutus}}, \bibinfo {author} {\bibfnamefont
  {O.}~\bibnamefont {Naaman}}, \bibinfo {author} {\bibfnamefont
  {M.}~\bibnamefont {Neeley}}, \bibinfo {author} {\bibfnamefont
  {C.}~\bibnamefont {Neill}}, \bibinfo {author} {\bibfnamefont {M.~Y.}\
  \bibnamefont {Niu}}, \bibinfo {author} {\bibfnamefont {E.}~\bibnamefont
  {Ostby}}, \bibinfo {author} {\bibfnamefont {A.}~\bibnamefont {Petukhov}},
  \bibinfo {author} {\bibfnamefont {J.~C.}\ \bibnamefont {Platt}}, \bibinfo
  {author} {\bibfnamefont {C.}~\bibnamefont {Quintana}}, \bibinfo {author}
  {\bibfnamefont {E.~G.}\ \bibnamefont {Rieffel}}, \bibinfo {author}
  {\bibfnamefont {P.}~\bibnamefont {Roushan}}, \bibinfo {author} {\bibfnamefont
  {N.~C.}\ \bibnamefont {Rubin}}, \bibinfo {author} {\bibfnamefont
  {D.}~\bibnamefont {Sank}}, \bibinfo {author} {\bibfnamefont {K.~J.}\
  \bibnamefont {Satzinger}}, \bibinfo {author} {\bibfnamefont {V.}~\bibnamefont
  {Smelyanskiy}}, \bibinfo {author} {\bibfnamefont {K.~J.}\ \bibnamefont
  {Sung}}, \bibinfo {author} {\bibfnamefont {M.~D.}\ \bibnamefont
  {Trevithick}}, \bibinfo {author} {\bibfnamefont {A.}~\bibnamefont
  {Vainsencher}}, \bibinfo {author} {\bibfnamefont {B.}~\bibnamefont
  {Villalonga}}, \bibinfo {author} {\bibfnamefont {T.}~\bibnamefont {White}},
  \bibinfo {author} {\bibfnamefont {Z.~J.}\ \bibnamefont {Yao}}, \bibinfo
  {author} {\bibfnamefont {P.}~\bibnamefont {Yeh}}, \bibinfo {author}
  {\bibfnamefont {A.}~\bibnamefont {Zalcman}}, \bibinfo {author} {\bibfnamefont
  {H.}~\bibnamefont {Neven}},\ and\ \bibinfo {author} {\bibfnamefont {J.~M.}\
  \bibnamefont {Martinis}},\ }\bibfield  {title} {\bibinfo {title} {Quantum
  supremacy using a programmable superconducting processor},\ }\href
  {https://doi.org/10.1038/s41586-019-1666-5} {\bibfield  {journal} {\bibinfo
  {journal} {Nature}\ }\textbf {\bibinfo {volume} {574}},\ \bibinfo {pages}
  {505} (\bibinfo {year} {2019})}\BibitemShut {NoStop}%
\bibitem [{\citenamefont {Zhong}\ \emph {et~al.}(2021)\citenamefont {Zhong},
  \citenamefont {Deng}, \citenamefont {Qin}, \citenamefont {Wang},
  \citenamefont {Chen}, \citenamefont {Peng}, \citenamefont {Luo},
  \citenamefont {Wu}, \citenamefont {Gong}, \citenamefont {Su}, \citenamefont
  {Hu}, \citenamefont {Hu}, \citenamefont {Yang}, \citenamefont {Zhang},
  \citenamefont {Li}, \citenamefont {Li}, \citenamefont {Jiang}, \citenamefont
  {Gan}, \citenamefont {Yang}, \citenamefont {You}, \citenamefont {Wang},
  \citenamefont {Li}, \citenamefont {Liu}, \citenamefont {Renema},
  \citenamefont {Lu},\ and\ \citenamefont {Pan}}]{Zhong:2021PRL}%
  \BibitemOpen
  \bibfield  {author} {\bibinfo {author} {\bibfnamefont {H.-S.}\ \bibnamefont
  {Zhong}}, \bibinfo {author} {\bibfnamefont {Y.-H.}\ \bibnamefont {Deng}},
  \bibinfo {author} {\bibfnamefont {J.}~\bibnamefont {Qin}}, \bibinfo {author}
  {\bibfnamefont {H.}~\bibnamefont {Wang}}, \bibinfo {author} {\bibfnamefont
  {M.-C.}\ \bibnamefont {Chen}}, \bibinfo {author} {\bibfnamefont {L.-C.}\
  \bibnamefont {Peng}}, \bibinfo {author} {\bibfnamefont {Y.-H.}\ \bibnamefont
  {Luo}}, \bibinfo {author} {\bibfnamefont {D.}~\bibnamefont {Wu}}, \bibinfo
  {author} {\bibfnamefont {S.-Q.}\ \bibnamefont {Gong}}, \bibinfo {author}
  {\bibfnamefont {H.}~\bibnamefont {Su}}, \bibinfo {author} {\bibfnamefont
  {Y.}~\bibnamefont {Hu}}, \bibinfo {author} {\bibfnamefont {P.}~\bibnamefont
  {Hu}}, \bibinfo {author} {\bibfnamefont {X.-Y.}\ \bibnamefont {Yang}},
  \bibinfo {author} {\bibfnamefont {W.-J.}\ \bibnamefont {Zhang}}, \bibinfo
  {author} {\bibfnamefont {H.}~\bibnamefont {Li}}, \bibinfo {author}
  {\bibfnamefont {Y.}~\bibnamefont {Li}}, \bibinfo {author} {\bibfnamefont
  {X.}~\bibnamefont {Jiang}}, \bibinfo {author} {\bibfnamefont
  {L.}~\bibnamefont {Gan}}, \bibinfo {author} {\bibfnamefont {G.}~\bibnamefont
  {Yang}}, \bibinfo {author} {\bibfnamefont {L.}~\bibnamefont {You}}, \bibinfo
  {author} {\bibfnamefont {Z.}~\bibnamefont {Wang}}, \bibinfo {author}
  {\bibfnamefont {L.}~\bibnamefont {Li}}, \bibinfo {author} {\bibfnamefont
  {N.-L.}\ \bibnamefont {Liu}}, \bibinfo {author} {\bibfnamefont {J.~J.}\
  \bibnamefont {Renema}}, \bibinfo {author} {\bibfnamefont {C.-Y.}\
  \bibnamefont {Lu}},\ and\ \bibinfo {author} {\bibfnamefont {J.-W.}\
  \bibnamefont {Pan}},\ }\bibfield  {title} {\bibinfo {title}
  {{Phase-Programmable Gaussian Boson Sampling Using Stimulated Squeezed
  Light}},\ }\href {https://doi.org/10.1103/PhysRevLett.127.180502} {\bibfield
  {journal} {\bibinfo  {journal} {Phys. Rev. Lett.}\ }\textbf {\bibinfo
  {volume} {127}},\ \bibinfo {pages} {180502} (\bibinfo {year}
  {2021})}\BibitemShut {NoStop}%
\bibitem [{\citenamefont {Pan}\ \emph {et~al.}(2022)\citenamefont {Pan},
  \citenamefont {Chen},\ and\ \citenamefont {Zhang}}]{pan_solving_2022}%
  \BibitemOpen
  \bibfield  {author} {\bibinfo {author} {\bibfnamefont {F.}~\bibnamefont
  {Pan}}, \bibinfo {author} {\bibfnamefont {K.}~\bibnamefont {Chen}},\ and\
  \bibinfo {author} {\bibfnamefont {P.}~\bibnamefont {Zhang}},\ }\bibfield
  {title} {\bibinfo {title} {{Solving the Sampling Problem of the Sycamore
  Quantum Circuits}},\ }\href {https://doi.org/10.1103/PhysRevLett.129.090502}
  {\bibfield  {journal} {\bibinfo  {journal} {Phys. Rev. Lett.}\ }\textbf
  {\bibinfo {volume} {129}},\ \bibinfo {pages} {090502} (\bibinfo {year}
  {2022})}\BibitemShut {NoStop}%
\bibitem [{\citenamefont {Clifford}\ and\ \citenamefont
  {Clifford}(2024)}]{clifford_faster_2024}%
  \BibitemOpen
  \bibfield  {author} {\bibinfo {author} {\bibfnamefont {P.}~\bibnamefont
  {Clifford}}\ and\ \bibinfo {author} {\bibfnamefont {R.}~\bibnamefont
  {Clifford}},\ }\bibfield  {title} {\bibinfo {title} {Faster classical boson
  sampling},\ }\href {https://doi.org/10.1088/1402-4896/ad4688} {\bibfield
  {journal} {\bibinfo  {journal} {Phys. Scr.}\ }\textbf {\bibinfo {volume}
  {99}},\ \bibinfo {pages} {065121} (\bibinfo {year} {2024})}\BibitemShut
  {NoStop}%
\bibitem [{\citenamefont {Clauser}\ \emph {et~al.}(1969)\citenamefont
  {Clauser}, \citenamefont {Horne}, \citenamefont {Shimony},\ and\
  \citenamefont {Holt}}]{Clauser:1969PRL}%
  \BibitemOpen
  \bibfield  {author} {\bibinfo {author} {\bibfnamefont {J.~F.}\ \bibnamefont
  {Clauser}}, \bibinfo {author} {\bibfnamefont {M.~A.}\ \bibnamefont {Horne}},
  \bibinfo {author} {\bibfnamefont {A.}~\bibnamefont {Shimony}},\ and\ \bibinfo
  {author} {\bibfnamefont {R.~A.}\ \bibnamefont {Holt}},\ }\bibfield  {title}
  {\bibinfo {title} {{Proposed Experiment to Test Local Hidden-Variable
  Theories}},\ }\href {https://doi.org/10.1103/PhysRevLett.23.880} {\bibfield
  {journal} {\bibinfo  {journal} {Phys. Rev. Lett.}\ }\textbf {\bibinfo
  {volume} {23}},\ \bibinfo {pages} {880} (\bibinfo {year} {1969})}\BibitemShut
  {NoStop}%
\bibitem [{\citenamefont {Steane}\ and\ \citenamefont {van
  Dam}(2000)}]{steane_physicists_2000}%
  \BibitemOpen
  \bibfield  {author} {\bibinfo {author} {\bibfnamefont {A.~M.}\ \bibnamefont
  {Steane}}\ and\ \bibinfo {author} {\bibfnamefont {W.}~\bibnamefont {van
  Dam}},\ }\bibfield  {title} {\bibinfo {title} {{Physicists Triumph at Guess
  My Number}},\ }\href {https://doi.org/10.1063/1.882963} {\bibfield  {journal}
  {\bibinfo  {journal} {Physics Today}\ }\textbf {\bibinfo {volume} {53}},\
  \bibinfo {pages} {35} (\bibinfo {year} {2000})}\BibitemShut {NoStop}%
\bibitem [{\citenamefont {Mermin}(1990)}]{Mermin:1990AJP}%
  \BibitemOpen
  \bibfield  {author} {\bibinfo {author} {\bibfnamefont {N.~D.}\ \bibnamefont
  {Mermin}},\ }\bibfield  {title} {\bibinfo {title} {Quantum mysteries
  revisited},\ }\href {https://doi.org/10.1119/1.16503} {\bibfield  {journal}
  {\bibinfo  {journal} {Am. J. Phys.}\ }\textbf {\bibinfo {volume} {58}},\
  \bibinfo {pages} {731} (\bibinfo {year} {1990})}\BibitemShut {NoStop}%
\bibitem [{\citenamefont {Pompili}\ \emph {et~al.}(2021)\citenamefont
  {Pompili}, \citenamefont {Hermans}, \citenamefont {Baier}, \citenamefont
  {Beukers}, \citenamefont {Humphreys}, \citenamefont {Schouten}, \citenamefont
  {Vermeulen}, \citenamefont {Tiggelman}, \citenamefont {dos Santos~Martins},
  \citenamefont {Dirkse}, \citenamefont {Wehner},\ and\ \citenamefont
  {Hanson}}]{pompili_realization_2021}%
  \BibitemOpen
  \bibfield  {author} {\bibinfo {author} {\bibfnamefont {M.}~\bibnamefont
  {Pompili}}, \bibinfo {author} {\bibfnamefont {S.~L.~N.}\ \bibnamefont
  {Hermans}}, \bibinfo {author} {\bibfnamefont {S.}~\bibnamefont {Baier}},
  \bibinfo {author} {\bibfnamefont {H.~K.~C.}\ \bibnamefont {Beukers}},
  \bibinfo {author} {\bibfnamefont {P.~C.}\ \bibnamefont {Humphreys}}, \bibinfo
  {author} {\bibfnamefont {R.~N.}\ \bibnamefont {Schouten}}, \bibinfo {author}
  {\bibfnamefont {R.~F.~L.}\ \bibnamefont {Vermeulen}}, \bibinfo {author}
  {\bibfnamefont {M.~J.}\ \bibnamefont {Tiggelman}}, \bibinfo {author}
  {\bibfnamefont {L.}~\bibnamefont {dos Santos~Martins}}, \bibinfo {author}
  {\bibfnamefont {B.}~\bibnamefont {Dirkse}}, \bibinfo {author} {\bibfnamefont
  {S.}~\bibnamefont {Wehner}},\ and\ \bibinfo {author} {\bibfnamefont
  {R.}~\bibnamefont {Hanson}},\ }\bibfield  {title} {\bibinfo {title}
  {Realization of a multinode quantum network of remote solid-state qubits},\
  }\href {https://doi.org/10.1126/science.abg1919} {\bibfield  {journal}
  {\bibinfo  {journal} {Science}\ }\textbf {\bibinfo {volume} {372}},\ \bibinfo
  {pages} {259} (\bibinfo {year} {2021})}\BibitemShut {NoStop}%
\bibitem [{\citenamefont {Pan}\ \emph {et~al.}(2000)\citenamefont {Pan},
  \citenamefont {Bouwmeester}, \citenamefont {Daniell}, \citenamefont
  {Weinfurter},\ and\ \citenamefont {Zeilinger}}]{Pan:2000NAT}%
  \BibitemOpen
  \bibfield  {author} {\bibinfo {author} {\bibfnamefont {J.-W.}\ \bibnamefont
  {Pan}}, \bibinfo {author} {\bibfnamefont {D.}~\bibnamefont {Bouwmeester}},
  \bibinfo {author} {\bibfnamefont {M.}~\bibnamefont {Daniell}}, \bibinfo
  {author} {\bibfnamefont {H.}~\bibnamefont {Weinfurter}},\ and\ \bibinfo
  {author} {\bibfnamefont {A.}~\bibnamefont {Zeilinger}},\ }\bibfield  {title}
  {\bibinfo {title} {{Experimental test of quantum nonlocality in three-photon
  Greenberger--Horne--Zeilinger entanglement}},\ }\href
  {https://doi.org/10.1038/35000514} {\bibfield  {journal} {\bibinfo  {journal}
  {Nature}\ }\textbf {\bibinfo {volume} {403}},\ \bibinfo {pages} {515}
  (\bibinfo {year} {2000})}\BibitemShut {NoStop}%
\bibitem [{\citenamefont {Cabello}\ \emph {et~al.}(2008)\citenamefont
  {Cabello}, \citenamefont {Rodr\'{\i}guez},\ and\ \citenamefont
  {Villanueva}}]{Cabello:2008PRL}%
  \BibitemOpen
  \bibfield  {author} {\bibinfo {author} {\bibfnamefont {A.}~\bibnamefont
  {Cabello}}, \bibinfo {author} {\bibfnamefont {D.}~\bibnamefont
  {Rodr\'{\i}guez}},\ and\ \bibinfo {author} {\bibfnamefont {I.}~\bibnamefont
  {Villanueva}},\ }\bibfield  {title} {\bibinfo {title} {{Necessary and
  Sufficient Detection Efficiency for the Mermin Inequalities}},\ }\href
  {https://doi.org/10.1103/PhysRevLett.101.120402} {\bibfield  {journal}
  {\bibinfo  {journal} {Phys. Rev. Lett.}\ }\textbf {\bibinfo {volume} {101}},\
  \bibinfo {pages} {120402} (\bibinfo {year} {2008})}\BibitemShut {NoStop}%
\bibitem [{\citenamefont {Cleve}\ \emph {et~al.}(2004)\citenamefont {Cleve},
  \citenamefont {{H{\o}yer}}, \citenamefont {Toner},\ and\ \citenamefont
  {Watrous}}]{cleve_consequences_2004}%
  \BibitemOpen
  \bibfield  {author} {\bibinfo {author} {\bibfnamefont {R.}~\bibnamefont
  {Cleve}}, \bibinfo {author} {\bibfnamefont {P.}~\bibnamefont {{H{\o}yer}}},
  \bibinfo {author} {\bibfnamefont {B.}~\bibnamefont {Toner}},\ and\ \bibinfo
  {author} {\bibfnamefont {J.}~\bibnamefont {Watrous}},\ }\bibfield  {title}
  {\bibinfo {title} {Consequences and limits of nonlocal strategies},\ }in\
  \href {https://doi.org/10.1109/CCC.2004.1313847} {\emph {\bibinfo {booktitle}
  {Proceedings. 19th IEEE Annual Conference on Computational Complexity,
  2004.}}}\ (\bibinfo {year} {2004})\ pp.\ \bibinfo {pages}
  {236--249}\BibitemShut {NoStop}%
\bibitem [{\citenamefont {Brassard}\ \emph {et~al.}(2005)\citenamefont
  {Brassard}, \citenamefont {Broadbent},\ and\ \citenamefont
  {Tapp}}]{brassard_quantum_2005}%
  \BibitemOpen
  \bibfield  {author} {\bibinfo {author} {\bibfnamefont {G.}~\bibnamefont
  {Brassard}}, \bibinfo {author} {\bibfnamefont {A.}~\bibnamefont
  {Broadbent}},\ and\ \bibinfo {author} {\bibfnamefont {A.}~\bibnamefont
  {Tapp}},\ }\bibfield  {title} {\bibinfo {title} {{Quantum
  Pseudo-Telepathy}},\ }\href {https://doi.org/10.1007/s10701-005-7353-4}
  {\bibfield  {journal} {\bibinfo  {journal} {Found. Phys.}\ }\textbf {\bibinfo
  {volume} {35}},\ \bibinfo {pages} {1877} (\bibinfo {year}
  {2005})}\BibitemShut {NoStop}%
\bibitem [{\citenamefont {Hensen}\ \emph {et~al.}(2015)\citenamefont {Hensen},
  \citenamefont {Bernien}, \citenamefont {Dr{\'e}au}, \citenamefont {Reiserer},
  \citenamefont {Kalb}, \citenamefont {Blok}, \citenamefont {Ruitenberg},
  \citenamefont {Vermeulen}, \citenamefont {Schouten}, \citenamefont
  {Abell{\'a}n}, \citenamefont {Amaya}, \citenamefont {Pruneri}, \citenamefont
  {Mitchell}, \citenamefont {Markham}, \citenamefont {Twitchen}, \citenamefont
  {Elkouss}, \citenamefont {Wehner}, \citenamefont {Taminiau},\ and\
  \citenamefont {Hanson}}]{hensen_loophole-free_2015}%
  \BibitemOpen
  \bibfield  {author} {\bibinfo {author} {\bibfnamefont {B.}~\bibnamefont
  {Hensen}}, \bibinfo {author} {\bibfnamefont {H.}~\bibnamefont {Bernien}},
  \bibinfo {author} {\bibfnamefont {A.~E.}\ \bibnamefont {Dr{\'e}au}}, \bibinfo
  {author} {\bibfnamefont {A.}~\bibnamefont {Reiserer}}, \bibinfo {author}
  {\bibfnamefont {N.}~\bibnamefont {Kalb}}, \bibinfo {author} {\bibfnamefont
  {M.~S.}\ \bibnamefont {Blok}}, \bibinfo {author} {\bibfnamefont
  {J.}~\bibnamefont {Ruitenberg}}, \bibinfo {author} {\bibfnamefont {R.~F.~L.}\
  \bibnamefont {Vermeulen}}, \bibinfo {author} {\bibfnamefont {R.~N.}\
  \bibnamefont {Schouten}}, \bibinfo {author} {\bibfnamefont {C.}~\bibnamefont
  {Abell{\'a}n}}, \bibinfo {author} {\bibfnamefont {W.}~\bibnamefont {Amaya}},
  \bibinfo {author} {\bibfnamefont {V.}~\bibnamefont {Pruneri}}, \bibinfo
  {author} {\bibfnamefont {M.~W.}\ \bibnamefont {Mitchell}}, \bibinfo {author}
  {\bibfnamefont {M.}~\bibnamefont {Markham}}, \bibinfo {author} {\bibfnamefont
  {D.~J.}\ \bibnamefont {Twitchen}}, \bibinfo {author} {\bibfnamefont
  {D.}~\bibnamefont {Elkouss}}, \bibinfo {author} {\bibfnamefont
  {S.}~\bibnamefont {Wehner}}, \bibinfo {author} {\bibfnamefont {T.~H.}\
  \bibnamefont {Taminiau}},\ and\ \bibinfo {author} {\bibfnamefont
  {R.}~\bibnamefont {Hanson}},\ }\bibfield  {title} {\bibinfo {title}
  {Loophole-free {B}ell inequality violation using electron spins separated by
  1.3 kilometres},\ }\href {https://doi.org/10.1038/nature15759} {\bibfield
  {journal} {\bibinfo  {journal} {Nature}\ }\textbf {\bibinfo {volume} {526}},\
  \bibinfo {pages} {682} (\bibinfo {year} {2015})}\BibitemShut {NoStop}%
\bibitem [{\citenamefont {Shalm}\ \emph {et~al.}(2015)\citenamefont {Shalm},
  \citenamefont {Meyer-Scott}, \citenamefont {Christensen}, \citenamefont
  {Bierhorst}, \citenamefont {Wayne}, \citenamefont {Stevens}, \citenamefont
  {Gerrits}, \citenamefont {Glancy}, \citenamefont {Hamel}, \citenamefont
  {Allman}, \citenamefont {Coakley}, \citenamefont {Dyer}, \citenamefont
  {Hodge}, \citenamefont {Lita}, \citenamefont {Verma}, \citenamefont
  {Lambrocco}, \citenamefont {Tortorici}, \citenamefont {Migdall},
  \citenamefont {Zhang}, \citenamefont {Kumor}, \citenamefont {Farr},
  \citenamefont {Marsili}, \citenamefont {Shaw}, \citenamefont {Stern},
  \citenamefont {Abell{\'a}n}, \citenamefont {Amaya}, \citenamefont {Pruneri},
  \citenamefont {Jennewein}, \citenamefont {Mitchell}, \citenamefont {Kwiat},
  \citenamefont {Bienfang}, \citenamefont {Mirin}, \citenamefont {Knill},\ and\
  \citenamefont {Nam}}]{shalm_strong_2015}%
  \BibitemOpen
  \bibfield  {author} {\bibinfo {author} {\bibfnamefont {L.~K.}\ \bibnamefont
  {Shalm}}, \bibinfo {author} {\bibfnamefont {E.}~\bibnamefont {Meyer-Scott}},
  \bibinfo {author} {\bibfnamefont {B.~G.}\ \bibnamefont {Christensen}},
  \bibinfo {author} {\bibfnamefont {P.}~\bibnamefont {Bierhorst}}, \bibinfo
  {author} {\bibfnamefont {M.~A.}\ \bibnamefont {Wayne}}, \bibinfo {author}
  {\bibfnamefont {M.~J.}\ \bibnamefont {Stevens}}, \bibinfo {author}
  {\bibfnamefont {T.}~\bibnamefont {Gerrits}}, \bibinfo {author} {\bibfnamefont
  {S.}~\bibnamefont {Glancy}}, \bibinfo {author} {\bibfnamefont {D.~R.}\
  \bibnamefont {Hamel}}, \bibinfo {author} {\bibfnamefont {M.~S.}\ \bibnamefont
  {Allman}}, \bibinfo {author} {\bibfnamefont {K.~J.}\ \bibnamefont {Coakley}},
  \bibinfo {author} {\bibfnamefont {S.~D.}\ \bibnamefont {Dyer}}, \bibinfo
  {author} {\bibfnamefont {C.}~\bibnamefont {Hodge}}, \bibinfo {author}
  {\bibfnamefont {A.~E.}\ \bibnamefont {Lita}}, \bibinfo {author}
  {\bibfnamefont {V.~B.}\ \bibnamefont {Verma}}, \bibinfo {author}
  {\bibfnamefont {C.}~\bibnamefont {Lambrocco}}, \bibinfo {author}
  {\bibfnamefont {E.}~\bibnamefont {Tortorici}}, \bibinfo {author}
  {\bibfnamefont {A.~L.}\ \bibnamefont {Migdall}}, \bibinfo {author}
  {\bibfnamefont {Y.}~\bibnamefont {Zhang}}, \bibinfo {author} {\bibfnamefont
  {D.~R.}\ \bibnamefont {Kumor}}, \bibinfo {author} {\bibfnamefont {W.~H.}\
  \bibnamefont {Farr}}, \bibinfo {author} {\bibfnamefont {F.}~\bibnamefont
  {Marsili}}, \bibinfo {author} {\bibfnamefont {M.~D.}\ \bibnamefont {Shaw}},
  \bibinfo {author} {\bibfnamefont {J.~A.}\ \bibnamefont {Stern}}, \bibinfo
  {author} {\bibfnamefont {C.}~\bibnamefont {Abell{\'a}n}}, \bibinfo {author}
  {\bibfnamefont {W.}~\bibnamefont {Amaya}}, \bibinfo {author} {\bibfnamefont
  {V.}~\bibnamefont {Pruneri}}, \bibinfo {author} {\bibfnamefont
  {T.}~\bibnamefont {Jennewein}}, \bibinfo {author} {\bibfnamefont {M.~W.}\
  \bibnamefont {Mitchell}}, \bibinfo {author} {\bibfnamefont {P.~G.}\
  \bibnamefont {Kwiat}}, \bibinfo {author} {\bibfnamefont {J.~C.}\ \bibnamefont
  {Bienfang}}, \bibinfo {author} {\bibfnamefont {R.~P.}\ \bibnamefont {Mirin}},
  \bibinfo {author} {\bibfnamefont {E.}~\bibnamefont {Knill}},\ and\ \bibinfo
  {author} {\bibfnamefont {S.~W.}\ \bibnamefont {Nam}},\ }\bibfield  {title}
  {\bibinfo {title} {{Strong Loophole-Free Test of Local Realism}},\ }\href
  {https://doi.org/10.1103/PhysRevLett.115.250402} {\bibfield  {journal}
  {\bibinfo  {journal} {Phys. Rev. Lett.}\ }\textbf {\bibinfo {volume} {115}},\
  \bibinfo {pages} {250402} (\bibinfo {year} {2015})}\BibitemShut {NoStop}%
\bibitem [{\citenamefont {Giustina}\ \emph {et~al.}(2015)\citenamefont
  {Giustina}, \citenamefont {Versteegh}, \citenamefont {Wengerowsky},
  \citenamefont {Handsteiner}, \citenamefont {Hochrainer}, \citenamefont
  {Phelan}, \citenamefont {Steinlechner}, \citenamefont {Kofler}, \citenamefont
  {Larsson}, \citenamefont {Abell\'an}, \citenamefont {Amaya}, \citenamefont
  {Pruneri}, \citenamefont {Mitchell}, \citenamefont {Beyer}, \citenamefont
  {Gerrits}, \citenamefont {Lita}, \citenamefont {Shalm}, \citenamefont {Nam},
  \citenamefont {Scheidl}, \citenamefont {Ursin}, \citenamefont {Wittmann},\
  and\ \citenamefont {Zeilinger}}]{giustina_significant_2015}%
  \BibitemOpen
  \bibfield  {author} {\bibinfo {author} {\bibfnamefont {M.}~\bibnamefont
  {Giustina}}, \bibinfo {author} {\bibfnamefont {M.~A.~M.}\ \bibnamefont
  {Versteegh}}, \bibinfo {author} {\bibfnamefont {S.}~\bibnamefont
  {Wengerowsky}}, \bibinfo {author} {\bibfnamefont {J.}~\bibnamefont
  {Handsteiner}}, \bibinfo {author} {\bibfnamefont {A.}~\bibnamefont
  {Hochrainer}}, \bibinfo {author} {\bibfnamefont {K.}~\bibnamefont {Phelan}},
  \bibinfo {author} {\bibfnamefont {F.}~\bibnamefont {Steinlechner}}, \bibinfo
  {author} {\bibfnamefont {J.}~\bibnamefont {Kofler}}, \bibinfo {author}
  {\bibfnamefont {J.-A.}\ \bibnamefont {Larsson}}, \bibinfo {author}
  {\bibfnamefont {C.}~\bibnamefont {Abell\'an}}, \bibinfo {author}
  {\bibfnamefont {W.}~\bibnamefont {Amaya}}, \bibinfo {author} {\bibfnamefont
  {V.}~\bibnamefont {Pruneri}}, \bibinfo {author} {\bibfnamefont {M.~W.}\
  \bibnamefont {Mitchell}}, \bibinfo {author} {\bibfnamefont {J.}~\bibnamefont
  {Beyer}}, \bibinfo {author} {\bibfnamefont {T.}~\bibnamefont {Gerrits}},
  \bibinfo {author} {\bibfnamefont {A.~E.}\ \bibnamefont {Lita}}, \bibinfo
  {author} {\bibfnamefont {L.~K.}\ \bibnamefont {Shalm}}, \bibinfo {author}
  {\bibfnamefont {S.~W.}\ \bibnamefont {Nam}}, \bibinfo {author} {\bibfnamefont
  {T.}~\bibnamefont {Scheidl}}, \bibinfo {author} {\bibfnamefont
  {R.}~\bibnamefont {Ursin}}, \bibinfo {author} {\bibfnamefont
  {B.}~\bibnamefont {Wittmann}},\ and\ \bibinfo {author} {\bibfnamefont
  {A.}~\bibnamefont {Zeilinger}},\ }\bibfield  {title} {\bibinfo {title}
  {{Significant-Loophole-Free Test of Bell's Theorem with Entangled Photons}},\
  }\href {https://doi.org/10.1103/PhysRevLett.115.250401} {\bibfield  {journal}
  {\bibinfo  {journal} {Phys. Rev. Lett.}\ }\textbf {\bibinfo {volume} {115}},\
  \bibinfo {pages} {250401} (\bibinfo {year} {2015})}\BibitemShut {NoStop}%
\bibitem [{\citenamefont {Rosenfeld}\ \emph {et~al.}(2017)\citenamefont
  {Rosenfeld}, \citenamefont {Burchardt}, \citenamefont {Garthoff},
  \citenamefont {Redeker}, \citenamefont {Ortegel}, \citenamefont {Rau},\ and\
  \citenamefont {Weinfurter}}]{rosenfeld_event-ready_2017}%
  \BibitemOpen
  \bibfield  {author} {\bibinfo {author} {\bibfnamefont {W.}~\bibnamefont
  {Rosenfeld}}, \bibinfo {author} {\bibfnamefont {D.}~\bibnamefont
  {Burchardt}}, \bibinfo {author} {\bibfnamefont {R.}~\bibnamefont {Garthoff}},
  \bibinfo {author} {\bibfnamefont {K.}~\bibnamefont {Redeker}}, \bibinfo
  {author} {\bibfnamefont {N.}~\bibnamefont {Ortegel}}, \bibinfo {author}
  {\bibfnamefont {M.}~\bibnamefont {Rau}},\ and\ \bibinfo {author}
  {\bibfnamefont {H.}~\bibnamefont {Weinfurter}},\ }\bibfield  {title}
  {\bibinfo {title} {{Event-Ready Bell Test Using Entangled Atoms
  Simultaneously Closing Detection and Locality Loopholes}},\ }\href
  {https://doi.org/10.1103/PhysRevLett.119.010402} {\bibfield  {journal}
  {\bibinfo  {journal} {Phys. Rev. Lett.}\ }\textbf {\bibinfo {volume} {119}},\
  \bibinfo {pages} {010402} (\bibinfo {year} {2017})}\BibitemShut {NoStop}%
\bibitem [{\citenamefont {Storz}\ \emph {et~al.}(2023)\citenamefont {Storz},
  \citenamefont {Sch{\"a}r}, \citenamefont {Kulikov}, \citenamefont {Magnard},
  \citenamefont {Kurpiers}, \citenamefont {L{\"u}tolf}, \citenamefont {Walter},
  \citenamefont {Copetudo}, \citenamefont {Reuer}, \citenamefont {Akin},
  \citenamefont {Besse}, \citenamefont {Gabureac}, \citenamefont {Norris},
  \citenamefont {Rosario}, \citenamefont {Martin}, \citenamefont {Martinez},
  \citenamefont {Amaya}, \citenamefont {Mitchell}, \citenamefont {Abellan},
  \citenamefont {Bancal}, \citenamefont {Sangouard}, \citenamefont {Royer},
  \citenamefont {Blais},\ and\ \citenamefont
  {Wallraff}}]{storz_loophole-free_2023}%
  \BibitemOpen
  \bibfield  {author} {\bibinfo {author} {\bibfnamefont {S.}~\bibnamefont
  {Storz}}, \bibinfo {author} {\bibfnamefont {J.}~\bibnamefont {Sch{\"a}r}},
  \bibinfo {author} {\bibfnamefont {A.}~\bibnamefont {Kulikov}}, \bibinfo
  {author} {\bibfnamefont {P.}~\bibnamefont {Magnard}}, \bibinfo {author}
  {\bibfnamefont {P.}~\bibnamefont {Kurpiers}}, \bibinfo {author}
  {\bibfnamefont {J.}~\bibnamefont {L{\"u}tolf}}, \bibinfo {author}
  {\bibfnamefont {T.}~\bibnamefont {Walter}}, \bibinfo {author} {\bibfnamefont
  {A.}~\bibnamefont {Copetudo}}, \bibinfo {author} {\bibfnamefont
  {K.}~\bibnamefont {Reuer}}, \bibinfo {author} {\bibfnamefont
  {A.}~\bibnamefont {Akin}}, \bibinfo {author} {\bibfnamefont {J.-C.}\
  \bibnamefont {Besse}}, \bibinfo {author} {\bibfnamefont {M.}~\bibnamefont
  {Gabureac}}, \bibinfo {author} {\bibfnamefont {G.~J.}\ \bibnamefont
  {Norris}}, \bibinfo {author} {\bibfnamefont {A.}~\bibnamefont {Rosario}},
  \bibinfo {author} {\bibfnamefont {F.}~\bibnamefont {Martin}}, \bibinfo
  {author} {\bibfnamefont {J.}~\bibnamefont {Martinez}}, \bibinfo {author}
  {\bibfnamefont {W.}~\bibnamefont {Amaya}}, \bibinfo {author} {\bibfnamefont
  {M.~W.}\ \bibnamefont {Mitchell}}, \bibinfo {author} {\bibfnamefont
  {C.}~\bibnamefont {Abellan}}, \bibinfo {author} {\bibfnamefont {J.-D.}\
  \bibnamefont {Bancal}}, \bibinfo {author} {\bibfnamefont {N.}~\bibnamefont
  {Sangouard}}, \bibinfo {author} {\bibfnamefont {B.}~\bibnamefont {Royer}},
  \bibinfo {author} {\bibfnamefont {A.}~\bibnamefont {Blais}},\ and\ \bibinfo
  {author} {\bibfnamefont {A.}~\bibnamefont {Wallraff}},\ }\bibfield  {title}
  {\bibinfo {title} {Loophole-free {B}ell inequality violation with
  superconducting circuits},\ }\href
  {https://doi.org/10.1038/s41586-023-05885-0} {\bibfield  {journal} {\bibinfo
  {journal} {Nature}\ }\textbf {\bibinfo {volume} {617}},\ \bibinfo {pages}
  {265} (\bibinfo {year} {2023})}\BibitemShut {NoStop}%
\bibitem [{\citenamefont {Cabello}(2001{\natexlab{a}})}]{Cabello:2001PRLa}%
  \BibitemOpen
  \bibfield  {author} {\bibinfo {author} {\bibfnamefont {A.}~\bibnamefont
  {Cabello}},\ }\bibfield  {title} {\bibinfo {title} {{Bell's Theorem without
  Inequalities and without Probabilities for Two Observers}},\ }\href
  {https://doi.org/10.1103/PhysRevLett.86.1911} {\bibfield  {journal} {\bibinfo
   {journal} {Phys. Rev. Lett.}\ }\textbf {\bibinfo {volume} {86}},\ \bibinfo
  {pages} {1911} (\bibinfo {year} {2001}{\natexlab{a}})}\BibitemShut {NoStop}%
\bibitem [{\citenamefont {Cabello}(2001{\natexlab{b}})}]{Cabello:2001PRLb}%
  \BibitemOpen
  \bibfield  {author} {\bibinfo {author} {\bibfnamefont {A.}~\bibnamefont
  {Cabello}},\ }\bibfield  {title} {\bibinfo {title} {{``All versus Nothing''
  Inseparability for Two Observers}},\ }\href
  {https://doi.org/10.1103/PhysRevLett.87.010403} {\bibfield  {journal}
  {\bibinfo  {journal} {Phys. Rev. Lett.}\ }\textbf {\bibinfo {volume} {87}},\
  \bibinfo {pages} {010403} (\bibinfo {year} {2001}{\natexlab{b}})}\BibitemShut
  {NoStop}%
\bibitem [{\citenamefont {Aravind}(2004)}]{Aravind:2004AJP}%
  \BibitemOpen
  \bibfield  {author} {\bibinfo {author} {\bibfnamefont {P.~K.}\ \bibnamefont
  {Aravind}},\ }\bibfield  {title} {\bibinfo {title} {Quantum mysteries
  revisited again},\ }\href {https://doi.org/10.1119/1.1773173} {\bibfield
  {journal} {\bibinfo  {journal} {Am. J. Phys.}\ }\textbf {\bibinfo {volume}
  {72}},\ \bibinfo {pages} {1303} (\bibinfo {year} {2004})}\BibitemShut
  {NoStop}%
\bibitem [{\citenamefont {Brassard}\ \emph {et~al.}(2004)\citenamefont
  {Brassard}, \citenamefont {Methot},\ and\ \citenamefont
  {Tapp}}]{brassard_minimum_2004}%
  \BibitemOpen
  \bibfield  {author} {\bibinfo {author} {\bibfnamefont {G.}~\bibnamefont
  {Brassard}}, \bibinfo {author} {\bibfnamefont {A.~A.}\ \bibnamefont
  {Methot}},\ and\ \bibinfo {author} {\bibfnamefont {A.}~\bibnamefont {Tapp}},\
  }\href {https://arxiv.org/abs/quant-ph/0412136} {\bibinfo {title} {Minimum
  entangled state dimension required for pseudo-telepathy}} (\bibinfo {year}
  {2004}),\ \bibinfo {note} {arXiv:quant-ph/0412136}\BibitemShut {NoStop}%
\bibitem [{\citenamefont {Xu}\ \emph {et~al.}(2022)\citenamefont {Xu},
  \citenamefont {Zhen}, \citenamefont {Yang}, \citenamefont {Cheng},
  \citenamefont {Ren}, \citenamefont {Chen}, \citenamefont {Wang},\ and\
  \citenamefont {Wang}}]{xu_experimental_2022}%
  \BibitemOpen
  \bibfield  {author} {\bibinfo {author} {\bibfnamefont {J.-M.}\ \bibnamefont
  {Xu}}, \bibinfo {author} {\bibfnamefont {Y.-Z.}\ \bibnamefont {Zhen}},
  \bibinfo {author} {\bibfnamefont {Y.-X.}\ \bibnamefont {Yang}}, \bibinfo
  {author} {\bibfnamefont {Z.-M.}\ \bibnamefont {Cheng}}, \bibinfo {author}
  {\bibfnamefont {Z.-C.}\ \bibnamefont {Ren}}, \bibinfo {author} {\bibfnamefont
  {K.}~\bibnamefont {Chen}}, \bibinfo {author} {\bibfnamefont {X.-L.}\
  \bibnamefont {Wang}},\ and\ \bibinfo {author} {\bibfnamefont {H.-T.}\
  \bibnamefont {Wang}},\ }\bibfield  {title} {\bibinfo {title} {{Experimental
  Demonstration of Quantum Pseudotelepathy}},\ }\href
  {https://doi.org/10.1103/PhysRevLett.129.050402} {\bibfield  {journal}
  {\bibinfo  {journal} {Phys. Rev. Lett.}\ }\textbf {\bibinfo {volume} {129}},\
  \bibinfo {pages} {050402} (\bibinfo {year} {2022})}\BibitemShut {NoStop}%
\bibitem [{\citenamefont {Weihs}\ \emph {et~al.}(1998)\citenamefont {Weihs},
  \citenamefont {Jennewein}, \citenamefont {Simon}, \citenamefont
  {Weinfurter},\ and\ \citenamefont {Zeilinger}}]{weihs_violation_1998}%
  \BibitemOpen
  \bibfield  {author} {\bibinfo {author} {\bibfnamefont {G.}~\bibnamefont
  {Weihs}}, \bibinfo {author} {\bibfnamefont {T.}~\bibnamefont {Jennewein}},
  \bibinfo {author} {\bibfnamefont {C.}~\bibnamefont {Simon}}, \bibinfo
  {author} {\bibfnamefont {H.}~\bibnamefont {Weinfurter}},\ and\ \bibinfo
  {author} {\bibfnamefont {A.}~\bibnamefont {Zeilinger}},\ }\bibfield  {title}
  {\bibinfo {title} {{Violation of Bell's Inequality under Strict Einstein
  Locality Conditions}},\ }\href {https://doi.org/10.1103/PhysRevLett.81.5039}
  {\bibfield  {journal} {\bibinfo  {journal} {Phys. Rev. Lett.}\ }\textbf
  {\bibinfo {volume} {81}},\ \bibinfo {pages} {5039} (\bibinfo {year}
  {1998})}\BibitemShut {NoStop}%
\bibitem [{\citenamefont {Zhao}\ \emph {et~al.}(2003)\citenamefont {Zhao},
  \citenamefont {Yang}, \citenamefont {Chen}, \citenamefont {Zhang},
  \citenamefont {\.{Z}ukowski},\ and\ \citenamefont
  {Pan}}]{zhao_experimental_2003}%
  \BibitemOpen
  \bibfield  {author} {\bibinfo {author} {\bibfnamefont {Z.}~\bibnamefont
  {Zhao}}, \bibinfo {author} {\bibfnamefont {T.}~\bibnamefont {Yang}}, \bibinfo
  {author} {\bibfnamefont {Y.-A.}\ \bibnamefont {Chen}}, \bibinfo {author}
  {\bibfnamefont {A.-N.}\ \bibnamefont {Zhang}}, \bibinfo {author}
  {\bibfnamefont {M.}~\bibnamefont {\.{Z}ukowski}},\ and\ \bibinfo {author}
  {\bibfnamefont {J.-W.}\ \bibnamefont {Pan}},\ }\bibfield  {title} {\bibinfo
  {title} {{Experimental Violation of Local Realism by Four-Photon
  Greenberger-Horne-Zeilinger Entanglement}},\ }\href
  {https://doi.org/10.1103/PhysRevLett.91.180401} {\bibfield  {journal}
  {\bibinfo  {journal} {Phys. Rev. Lett.}\ }\textbf {\bibinfo {volume} {91}},\
  \bibinfo {pages} {180401} (\bibinfo {year} {2003})}\BibitemShut {NoStop}%
\bibitem [{\citenamefont {Huang}\ \emph {et~al.}(2003)\citenamefont {Huang},
  \citenamefont {Li}, \citenamefont {Zhang}, \citenamefont {Pan},\ and\
  \citenamefont {Guo}}]{huang_experimental_2003}%
  \BibitemOpen
  \bibfield  {author} {\bibinfo {author} {\bibfnamefont {Y.-F.}\ \bibnamefont
  {Huang}}, \bibinfo {author} {\bibfnamefont {C.-F.}\ \bibnamefont {Li}},
  \bibinfo {author} {\bibfnamefont {Y.-S.}\ \bibnamefont {Zhang}}, \bibinfo
  {author} {\bibfnamefont {J.-W.}\ \bibnamefont {Pan}},\ and\ \bibinfo {author}
  {\bibfnamefont {G.-C.}\ \bibnamefont {Guo}},\ }\bibfield  {title} {\bibinfo
  {title} {{Experimental Test of the Kochen-Specker Theorem with Single
  Photons}},\ }\href {https://doi.org/10.1103/PhysRevLett.90.250401} {\bibfield
   {journal} {\bibinfo  {journal} {Phys. Rev. Lett.}\ }\textbf {\bibinfo
  {volume} {90}},\ \bibinfo {pages} {250401} (\bibinfo {year}
  {2003})}\BibitemShut {NoStop}%
\bibitem [{\citenamefont {Pomarico}\ \emph {et~al.}(2011)\citenamefont
  {Pomarico}, \citenamefont {Bancal}, \citenamefont {Sanguinetti},
  \citenamefont {Rochdi},\ and\ \citenamefont {Gisin}}]{pomarico_various_2011}%
  \BibitemOpen
  \bibfield  {author} {\bibinfo {author} {\bibfnamefont {E.}~\bibnamefont
  {Pomarico}}, \bibinfo {author} {\bibfnamefont {J.-D.}\ \bibnamefont
  {Bancal}}, \bibinfo {author} {\bibfnamefont {B.}~\bibnamefont {Sanguinetti}},
  \bibinfo {author} {\bibfnamefont {A.}~\bibnamefont {Rochdi}},\ and\ \bibinfo
  {author} {\bibfnamefont {N.}~\bibnamefont {Gisin}},\ }\bibfield  {title}
  {\bibinfo {title} {Various quantum nonlocality tests with a commercial
  two-photon entanglement source},\ }\href
  {https://doi.org/10.1103/PhysRevA.83.052104} {\bibfield  {journal} {\bibinfo
  {journal} {Phys. Rev. A}\ }\textbf {\bibinfo {volume} {83}},\ \bibinfo
  {pages} {052104} (\bibinfo {year} {2011})}\BibitemShut {NoStop}%
\bibitem [{\citenamefont {Aolita}\ \emph {et~al.}(2012)\citenamefont {Aolita},
  \citenamefont {Gallego}, \citenamefont {Ac\'{\i}n}, \citenamefont {Chiuri},
  \citenamefont {Vallone}, \citenamefont {Mataloni},\ and\ \citenamefont
  {Cabello}}]{Aolita:2012PRA}%
  \BibitemOpen
  \bibfield  {author} {\bibinfo {author} {\bibfnamefont {L.}~\bibnamefont
  {Aolita}}, \bibinfo {author} {\bibfnamefont {R.}~\bibnamefont {Gallego}},
  \bibinfo {author} {\bibfnamefont {A.}~\bibnamefont {Ac\'{\i}n}}, \bibinfo
  {author} {\bibfnamefont {A.}~\bibnamefont {Chiuri}}, \bibinfo {author}
  {\bibfnamefont {G.}~\bibnamefont {Vallone}}, \bibinfo {author} {\bibfnamefont
  {P.}~\bibnamefont {Mataloni}},\ and\ \bibinfo {author} {\bibfnamefont
  {A.}~\bibnamefont {Cabello}},\ }\bibfield  {title} {\bibinfo {title} {Fully
  nonlocal quantum correlations},\ }\href
  {https://doi.org/10.1103/PhysRevA.85.032107} {\bibfield  {journal} {\bibinfo
  {journal} {Phys. Rev. A}\ }\textbf {\bibinfo {volume} {85}},\ \bibinfo
  {pages} {032107} (\bibinfo {year} {2012})}\BibitemShut {NoStop}%
\bibitem [{\citenamefont {Christensen}\ \emph {et~al.}(2015)\citenamefont
  {Christensen}, \citenamefont {Liang}, \citenamefont {Brunner}, \citenamefont
  {Gisin},\ and\ \citenamefont {Kwiat}}]{Christensen:2015PRX}%
  \BibitemOpen
  \bibfield  {author} {\bibinfo {author} {\bibfnamefont {B.~G.}\ \bibnamefont
  {Christensen}}, \bibinfo {author} {\bibfnamefont {Y.-C.}\ \bibnamefont
  {Liang}}, \bibinfo {author} {\bibfnamefont {N.}~\bibnamefont {Brunner}},
  \bibinfo {author} {\bibfnamefont {N.}~\bibnamefont {Gisin}},\ and\ \bibinfo
  {author} {\bibfnamefont {P.~G.}\ \bibnamefont {Kwiat}},\ }\bibfield  {title}
  {\bibinfo {title} {{Exploring the Limits of Quantum Nonlocality with
  Entangled Photons}},\ }\href {https://doi.org/10.1103/PhysRevX.5.041052}
  {\bibfield  {journal} {\bibinfo  {journal} {Phys. Rev. X}\ }\textbf {\bibinfo
  {volume} {5}},\ \bibinfo {pages} {041052} (\bibinfo {year}
  {2015})}\BibitemShut {NoStop}%
\bibitem [{\citenamefont {Hu}\ \emph {et~al.}(2016)\citenamefont {Hu},
  \citenamefont {Chen}, \citenamefont {Liu}, \citenamefont {Guo}, \citenamefont
  {Huang}, \citenamefont {Zhou}, \citenamefont {Han}, \citenamefont {Li},\ and\
  \citenamefont {Guo}}]{hu_experimental_2016}%
  \BibitemOpen
  \bibfield  {author} {\bibinfo {author} {\bibfnamefont {X.-M.}\ \bibnamefont
  {Hu}}, \bibinfo {author} {\bibfnamefont {J.-S.}\ \bibnamefont {Chen}},
  \bibinfo {author} {\bibfnamefont {B.-H.}\ \bibnamefont {Liu}}, \bibinfo
  {author} {\bibfnamefont {Y.}~\bibnamefont {Guo}}, \bibinfo {author}
  {\bibfnamefont {Y.-F.}\ \bibnamefont {Huang}}, \bibinfo {author}
  {\bibfnamefont {Z.-Q.}\ \bibnamefont {Zhou}}, \bibinfo {author}
  {\bibfnamefont {Y.-J.}\ \bibnamefont {Han}}, \bibinfo {author} {\bibfnamefont
  {C.-F.}\ \bibnamefont {Li}},\ and\ \bibinfo {author} {\bibfnamefont {G.-C.}\
  \bibnamefont {Guo}},\ }\bibfield  {title} {\bibinfo {title} {{Experimental
  Test of Compatibility-Loophole-Free Contextuality with Spatially Separated
  Entangled Qutrits}},\ }\href {https://doi.org/10.1103/PhysRevLett.117.170403}
  {\bibfield  {journal} {\bibinfo  {journal} {Phys. Rev. Lett.}\ }\textbf
  {\bibinfo {volume} {117}},\ \bibinfo {pages} {170403} (\bibinfo {year}
  {2016})}\BibitemShut {NoStop}%
\bibitem [{\citenamefont {Qu}\ \emph {et~al.}(2021)\citenamefont {Qu},
  \citenamefont {Wang}, \citenamefont {Xiao}, \citenamefont {Zhan},\ and\
  \citenamefont {Xue}}]{qu_state-independent_2021}%
  \BibitemOpen
  \bibfield  {author} {\bibinfo {author} {\bibfnamefont {D.}~\bibnamefont
  {Qu}}, \bibinfo {author} {\bibfnamefont {K.}~\bibnamefont {Wang}}, \bibinfo
  {author} {\bibfnamefont {L.}~\bibnamefont {Xiao}}, \bibinfo {author}
  {\bibfnamefont {X.}~\bibnamefont {Zhan}},\ and\ \bibinfo {author}
  {\bibfnamefont {P.}~\bibnamefont {Xue}},\ }\bibfield  {title} {\bibinfo
  {title} {State-independent test of quantum contextuality with either single
  photons or coherent light},\ }\href
  {https://doi.org/10.1038/s41534-021-00492-1} {\bibfield  {journal} {\bibinfo
  {journal} {npj Quantum Inf.}\ }\textbf {\bibinfo {volume} {7}},\ \bibinfo
  {pages} {154} (\bibinfo {year} {2021})}\BibitemShut {NoStop}%
\bibitem [{\citenamefont {Rowe}\ \emph {et~al.}(2001)\citenamefont {Rowe},
  \citenamefont {Kielpinski}, \citenamefont {Meyer}, \citenamefont {Sackett},
  \citenamefont {Itano}, \citenamefont {Monroe},\ and\ \citenamefont
  {Wineland}}]{rowe_experimental_2001}%
  \BibitemOpen
  \bibfield  {author} {\bibinfo {author} {\bibfnamefont {M.~A.}\ \bibnamefont
  {Rowe}}, \bibinfo {author} {\bibfnamefont {D.}~\bibnamefont {Kielpinski}},
  \bibinfo {author} {\bibfnamefont {V.}~\bibnamefont {Meyer}}, \bibinfo
  {author} {\bibfnamefont {C.~A.}\ \bibnamefont {Sackett}}, \bibinfo {author}
  {\bibfnamefont {W.~M.}\ \bibnamefont {Itano}}, \bibinfo {author}
  {\bibfnamefont {C.}~\bibnamefont {Monroe}},\ and\ \bibinfo {author}
  {\bibfnamefont {D.~J.}\ \bibnamefont {Wineland}},\ }\bibfield  {title}
  {\bibinfo {title} {Experimental violation of a {B}ell's inequality with
  efficient detection},\ }\href {https://doi.org/10.1038/35057215} {\bibfield
  {journal} {\bibinfo  {journal} {Nature}\ }\textbf {\bibinfo {volume} {409}},\
  \bibinfo {pages} {791} (\bibinfo {year} {2001})}\BibitemShut {NoStop}%
\bibitem [{\citenamefont {Matsukevich}\ \emph {et~al.}(2008)\citenamefont
  {Matsukevich}, \citenamefont {Maunz}, \citenamefont {Moehring}, \citenamefont
  {Olmschenk},\ and\ \citenamefont {Monroe}}]{matsukevich_bell_2008}%
  \BibitemOpen
  \bibfield  {author} {\bibinfo {author} {\bibfnamefont {D.~N.}\ \bibnamefont
  {Matsukevich}}, \bibinfo {author} {\bibfnamefont {P.}~\bibnamefont {Maunz}},
  \bibinfo {author} {\bibfnamefont {D.~L.}\ \bibnamefont {Moehring}}, \bibinfo
  {author} {\bibfnamefont {S.}~\bibnamefont {Olmschenk}},\ and\ \bibinfo
  {author} {\bibfnamefont {C.}~\bibnamefont {Monroe}},\ }\bibfield  {title}
  {\bibinfo {title} {{Bell Inequality Violation with Two Remote Atomic
  Qubits}},\ }\href {https://doi.org/10.1103/PhysRevLett.100.150404} {\bibfield
   {journal} {\bibinfo  {journal} {Phys. Rev. Lett.}\ }\textbf {\bibinfo
  {volume} {100}},\ \bibinfo {pages} {150404} (\bibinfo {year}
  {2008})}\BibitemShut {NoStop}%
\bibitem [{\citenamefont {Kirchmair}\ \emph {et~al.}(2009)\citenamefont
  {Kirchmair}, \citenamefont {Zähringer}, \citenamefont {Gerritsma},
  \citenamefont {Kleinmann}, \citenamefont {Gühne}, \citenamefont {Cabello},
  \citenamefont {Blatt},\ and\ \citenamefont
  {Roos}}]{kirchmair_state-independent_2009}%
  \BibitemOpen
  \bibfield  {author} {\bibinfo {author} {\bibfnamefont {G.}~\bibnamefont
  {Kirchmair}}, \bibinfo {author} {\bibfnamefont {F.}~\bibnamefont
  {Zähringer}}, \bibinfo {author} {\bibfnamefont {R.}~\bibnamefont
  {Gerritsma}}, \bibinfo {author} {\bibfnamefont {M.}~\bibnamefont
  {Kleinmann}}, \bibinfo {author} {\bibfnamefont {O.}~\bibnamefont {Gühne}},
  \bibinfo {author} {\bibfnamefont {A.}~\bibnamefont {Cabello}}, \bibinfo
  {author} {\bibfnamefont {R.}~\bibnamefont {Blatt}},\ and\ \bibinfo {author}
  {\bibfnamefont {C.~F.}\ \bibnamefont {Roos}},\ }\bibfield  {title} {\bibinfo
  {title} {State-independent experimental test of quantum contextuality},\
  }\href {https://doi.org/10.1038/nature08172} {\bibfield  {journal} {\bibinfo
  {journal} {Nature}\ }\textbf {\bibinfo {volume} {460}},\ \bibinfo {pages}
  {494} (\bibinfo {year} {2009})}\BibitemShut {NoStop}%
\bibitem [{\citenamefont {Tan}\ \emph {et~al.}(2017)\citenamefont {Tan},
  \citenamefont {Wan}, \citenamefont {Erickson}, \citenamefont {Bierhorst},
  \citenamefont {Kienzler}, \citenamefont {Glancy}, \citenamefont {Knill},
  \citenamefont {Leibfried},\ and\ \citenamefont
  {Wineland}}]{tan_chained_2017}%
  \BibitemOpen
  \bibfield  {author} {\bibinfo {author} {\bibfnamefont {T.~R.}\ \bibnamefont
  {Tan}}, \bibinfo {author} {\bibfnamefont {Y.}~\bibnamefont {Wan}}, \bibinfo
  {author} {\bibfnamefont {S.}~\bibnamefont {Erickson}}, \bibinfo {author}
  {\bibfnamefont {P.}~\bibnamefont {Bierhorst}}, \bibinfo {author}
  {\bibfnamefont {D.}~\bibnamefont {Kienzler}}, \bibinfo {author}
  {\bibfnamefont {S.}~\bibnamefont {Glancy}}, \bibinfo {author} {\bibfnamefont
  {E.}~\bibnamefont {Knill}}, \bibinfo {author} {\bibfnamefont
  {D.}~\bibnamefont {Leibfried}},\ and\ \bibinfo {author} {\bibfnamefont
  {D.~J.}\ \bibnamefont {Wineland}},\ }\bibfield  {title} {\bibinfo {title}
  {{Chained Bell Inequality Experiment with High-Efficiency Measurements}},\
  }\href {https://doi.org/10.1103/PhysRevLett.118.130403} {\bibfield  {journal}
  {\bibinfo  {journal} {Phys. Rev. Lett.}\ }\textbf {\bibinfo {volume} {118}},\
  \bibinfo {pages} {130403} (\bibinfo {year} {2017})}\BibitemShut {NoStop}%
\bibitem [{\citenamefont {Stephenson}\ \emph {et~al.}(2020)\citenamefont
  {Stephenson}, \citenamefont {Nadlinger}, \citenamefont {Nichol},
  \citenamefont {An}, \citenamefont {Drmota}, \citenamefont {Ballance},
  \citenamefont {Thirumalai}, \citenamefont {Goodwin}, \citenamefont {Lucas},\
  and\ \citenamefont {Ballance}}]{stephenson_high-rate_2020}%
  \BibitemOpen
  \bibfield  {author} {\bibinfo {author} {\bibfnamefont {L.~J.}\ \bibnamefont
  {Stephenson}}, \bibinfo {author} {\bibfnamefont {D.~P.}\ \bibnamefont
  {Nadlinger}}, \bibinfo {author} {\bibfnamefont {B.~C.}\ \bibnamefont
  {Nichol}}, \bibinfo {author} {\bibfnamefont {S.}~\bibnamefont {An}}, \bibinfo
  {author} {\bibfnamefont {P.}~\bibnamefont {Drmota}}, \bibinfo {author}
  {\bibfnamefont {T.~G.}\ \bibnamefont {Ballance}}, \bibinfo {author}
  {\bibfnamefont {K.}~\bibnamefont {Thirumalai}}, \bibinfo {author}
  {\bibfnamefont {J.~F.}\ \bibnamefont {Goodwin}}, \bibinfo {author}
  {\bibfnamefont {D.~M.}\ \bibnamefont {Lucas}},\ and\ \bibinfo {author}
  {\bibfnamefont {C.~J.}\ \bibnamefont {Ballance}},\ }\bibfield  {title}
  {\bibinfo {title} {{High-Rate, High-Fidelity Entanglement of Qubits Across an
  Elementary Quantum Network}},\ }\href
  {https://doi.org/10.1103/PhysRevLett.124.110501} {\bibfield  {journal}
  {\bibinfo  {journal} {Phys. Rev. Lett.}\ }\textbf {\bibinfo {volume} {124}},\
  \bibinfo {pages} {110501} (\bibinfo {year} {2020})}\BibitemShut {NoStop}%
\bibitem [{\citenamefont {Braunstein}\ and\ \citenamefont
  {Caves}(1990)}]{braunstein_wringing_1990}%
  \BibitemOpen
  \bibfield  {author} {\bibinfo {author} {\bibfnamefont {S.~L.}\ \bibnamefont
  {Braunstein}}\ and\ \bibinfo {author} {\bibfnamefont {C.~M.}\ \bibnamefont
  {Caves}},\ }\bibfield  {title} {\bibinfo {title} {Wringing out better {B}ell
  inequalities},\ }\href
  {https://doi.org/https://doi.org/10.1016/0003-4916(90)90339-P} {\bibfield
  {journal} {\bibinfo  {journal} {Ann. Phys.}\ }\textbf {\bibinfo {volume}
  {202}},\ \bibinfo {pages} {22} (\bibinfo {year} {1990})}\BibitemShut
  {NoStop}%
\bibitem [{\citenamefont {Elitzur}\ \emph {et~al.}(1992)\citenamefont
  {Elitzur}, \citenamefont {Popescu},\ and\ \citenamefont
  {Rohrlich}}]{Elitzur:1992PLA}%
  \BibitemOpen
  \bibfield  {author} {\bibinfo {author} {\bibfnamefont {A.~C.}\ \bibnamefont
  {Elitzur}}, \bibinfo {author} {\bibfnamefont {S.}~\bibnamefont {Popescu}},\
  and\ \bibinfo {author} {\bibfnamefont {D.}~\bibnamefont {Rohrlich}},\
  }\bibfield  {title} {\bibinfo {title} {Quantum nonlocality for each pair in
  an ensemble},\ }\href
  {https://doi.org/https://doi.org/10.1016/0375-9601(92)90952-I} {\bibfield
  {journal} {\bibinfo  {journal} {Phys. Lett. A}\ }\textbf {\bibinfo {volume}
  {162}},\ \bibinfo {pages} {25} (\bibinfo {year} {1992})}\BibitemShut
  {NoStop}%
\bibitem [{\citenamefont {\v{C}aslav Brukner}\ \emph
  {et~al.}(2006)\citenamefont {\v{C}aslav Brukner}, \citenamefont
  {Paunkovi\'{c}}, \citenamefont {Rudolph},\ and\ \citenamefont
  {Vedral}}]{brukner_entanglement-assisted_2006}%
  \BibitemOpen
  \bibfield  {author} {\bibinfo {author} {\bibnamefont {\v{C}aslav Brukner}},
  \bibinfo {author} {\bibfnamefont {N.}~\bibnamefont {Paunkovi\'{c}}}, \bibinfo
  {author} {\bibfnamefont {T.}~\bibnamefont {Rudolph}},\ and\ \bibinfo {author}
  {\bibfnamefont {V.}~\bibnamefont {Vedral}},\ }\bibfield  {title} {\bibinfo
  {title} {Entanglement-assisted orientation in space},\ }\href
  {https://doi.org/10.1142/S0219749906001839} {\bibfield  {journal} {\bibinfo
  {journal} {Int. J. Quantum Inf.}\ }\textbf {\bibinfo {volume} {04}},\
  \bibinfo {pages} {365} (\bibinfo {year} {2006})}\BibitemShut {NoStop}%
\bibitem [{\citenamefont {Mironowicz}(2023)}]{mironowicz_entangled_2023}%
  \BibitemOpen
  \bibfield  {author} {\bibinfo {author} {\bibfnamefont {P.}~\bibnamefont
  {Mironowicz}},\ }\bibfield  {title} {\bibinfo {title} {Entangled rendezvous:
  a possible application of bell non-locality for mobile agents on networks},\
  }\href@noop {} {\bibfield  {journal} {\bibinfo  {journal} {New J. Phys.}\
  }\textbf {\bibinfo {volume} {25}},\ \bibinfo {pages} {013023} (\bibinfo
  {year} {2023})}\BibitemShut {NoStop}%
\bibitem [{\citenamefont {Viola}\ and\ \citenamefont
  {Mironowicz}(2024)}]{viola_quantum_2024}%
  \BibitemOpen
  \bibfield  {author} {\bibinfo {author} {\bibfnamefont {G.}~\bibnamefont
  {Viola}}\ and\ \bibinfo {author} {\bibfnamefont {P.}~\bibnamefont
  {Mironowicz}},\ }\bibfield  {title} {\bibinfo {title} {Quantum strategies for
  rendezvous and domination tasks on graphs with mobile agents},\ }\href
  {https://doi.org/10.1103/PhysRevA.109.042201} {\bibfield  {journal} {\bibinfo
   {journal} {Phys. Rev. A}\ }\textbf {\bibinfo {volume} {109}},\ \bibinfo
  {pages} {042201} (\bibinfo {year} {2024})}\BibitemShut {NoStop}%
\bibitem [{\citenamefont {Tucker}\ \emph {et~al.}(2024)\citenamefont {Tucker},
  \citenamefont {Strange}, \citenamefont {Mironowicz},\ and\ \citenamefont
  {Quintanilla}}]{tucker_quantum-assisted_2024}%
  \BibitemOpen
  \bibfield  {author} {\bibinfo {author} {\bibfnamefont {J.}~\bibnamefont
  {Tucker}}, \bibinfo {author} {\bibfnamefont {P.}~\bibnamefont {Strange}},
  \bibinfo {author} {\bibfnamefont {P.}~\bibnamefont {Mironowicz}},\ and\
  \bibinfo {author} {\bibfnamefont {J.}~\bibnamefont {Quintanilla}},\ }\href
  {https://arxiv.org/abs/2405.14951} {\bibinfo {title} {Quantum-assisted
  rendezvous on graphs: Explicit algorithms and quantum computer simulations}}
  (\bibinfo {year} {2024}),\ \bibinfo {note} {arXiv:2405.14951}\BibitemShut
  {NoStop}%
\bibitem [{\citenamefont {Drmota}\ \emph {et~al.}(2023)\citenamefont {Drmota},
  \citenamefont {Main}, \citenamefont {Nadlinger}, \citenamefont {Nichol},
  \citenamefont {Weber}, \citenamefont {Ainley}, \citenamefont {Agrawal},
  \citenamefont {Srinivas}, \citenamefont {Araneda}, \citenamefont {Ballance},\
  and\ \citenamefont {Lucas}}]{drmota_robust_2023}%
  \BibitemOpen
  \bibfield  {author} {\bibinfo {author} {\bibfnamefont {P.}~\bibnamefont
  {Drmota}}, \bibinfo {author} {\bibfnamefont {D.}~\bibnamefont {Main}},
  \bibinfo {author} {\bibfnamefont {D.~P.}\ \bibnamefont {Nadlinger}}, \bibinfo
  {author} {\bibfnamefont {B.~C.}\ \bibnamefont {Nichol}}, \bibinfo {author}
  {\bibfnamefont {M.~A.}\ \bibnamefont {Weber}}, \bibinfo {author}
  {\bibfnamefont {E.~M.}\ \bibnamefont {Ainley}}, \bibinfo {author}
  {\bibfnamefont {A.}~\bibnamefont {Agrawal}}, \bibinfo {author} {\bibfnamefont
  {R.}~\bibnamefont {Srinivas}}, \bibinfo {author} {\bibfnamefont
  {G.}~\bibnamefont {Araneda}}, \bibinfo {author} {\bibfnamefont {C.~J.}\
  \bibnamefont {Ballance}},\ and\ \bibinfo {author} {\bibfnamefont {D.~M.}\
  \bibnamefont {Lucas}},\ }\bibfield  {title} {\bibinfo {title} {{Robust
  Quantum Memory in a Trapped-Ion Quantum Network Node}},\ }\href
  {https://doi.org/10.1103/PhysRevLett.130.090803} {\bibfield  {journal}
  {\bibinfo  {journal} {Phys. Rev. Lett.}\ }\textbf {\bibinfo {volume} {130}},\
  \bibinfo {pages} {090803} (\bibinfo {year} {2023})}\BibitemShut {NoStop}%
\bibitem [{\citenamefont {Main}\ \emph {et~al.}(2024)\citenamefont {Main} \emph
  {et~al.}}]{mixed_species_entanglement}%
  \BibitemOpen
  \bibfield  {author} {\bibinfo {author} {\bibfnamefont {D.}~\bibnamefont
  {Main}} \emph {et~al.},\ }\href@noop {} {\bibinfo {title} {Manuscript in
  preparation}} (\bibinfo {year} {2024})\BibitemShut {NoStop}%
\bibitem [{\citenamefont {Bourdeauducq}\ \emph {et~al.}(2021)\citenamefont
  {Bourdeauducq} \emph {et~al.}}]{ARTIQ}%
  \BibitemOpen
  \bibfield  {author} {\bibinfo {author} {\bibfnamefont {S.}~\bibnamefont
  {Bourdeauducq}} \emph {et~al.},\ }\href
  {https://doi.org/10.5281/zenodo.1492176} {\bibinfo {title} {{m-labs/artiq:
  6.0 (Version 6.0)}}} (\bibinfo {year} {2021})\BibitemShut {NoStop}%
\bibitem [{\citenamefont {Cabello}\ \emph {et~al.}(2014)\citenamefont
  {Cabello}, \citenamefont {Severini},\ and\ \citenamefont
  {Winter}}]{CSWPRL2014}%
  \BibitemOpen
  \bibfield  {author} {\bibinfo {author} {\bibfnamefont {A.}~\bibnamefont
  {Cabello}}, \bibinfo {author} {\bibfnamefont {S.}~\bibnamefont {Severini}},\
  and\ \bibinfo {author} {\bibfnamefont {A.}~\bibnamefont {Winter}},\
  }\bibfield  {title} {\bibinfo {title} {{Graph-Theoretic Approach to Quantum
  Correlations}},\ }\href {https://doi.org/10.1103/PhysRevLett.112.040401}
  {\bibfield  {journal} {\bibinfo  {journal} {Phys. Rev. Lett.}\ }\textbf
  {\bibinfo {volume} {112}},\ \bibinfo {pages} {040401} (\bibinfo {year}
  {2014})}\BibitemShut {NoStop}%
\bibitem [{\citenamefont {Ara\'ujo}\ \emph {et~al.}(2013)\citenamefont
  {Ara\'ujo}, \citenamefont {Quintino}, \citenamefont {Budroni}, \citenamefont
  {Cunha},\ and\ \citenamefont {Cabello}}]{Araujo:2013PRA}%
  \BibitemOpen
  \bibfield  {author} {\bibinfo {author} {\bibfnamefont {M.}~\bibnamefont
  {Ara\'ujo}}, \bibinfo {author} {\bibfnamefont {M.~T.}\ \bibnamefont
  {Quintino}}, \bibinfo {author} {\bibfnamefont {C.}~\bibnamefont {Budroni}},
  \bibinfo {author} {\bibfnamefont {M.~T.}\ \bibnamefont {Cunha}},\ and\
  \bibinfo {author} {\bibfnamefont {A.}~\bibnamefont {Cabello}},\ }\bibfield
  {title} {\bibinfo {title} {All noncontextuality inequalities for the
  $n$-cycle scenario},\ }\href {https://doi.org/10.1103/PhysRevA.88.022118}
  {\bibfield  {journal} {\bibinfo  {journal} {Phys. Rev. A}\ }\textbf {\bibinfo
  {volume} {88}},\ \bibinfo {pages} {022118} (\bibinfo {year}
  {2013})}\BibitemShut {NoStop}%
\bibitem [{\citenamefont {Bharti}\ \emph {et~al.}(2022)\citenamefont {Bharti},
  \citenamefont {Ray}, \citenamefont {Xu}, \citenamefont {Hayashi},
  \citenamefont {Kwek},\ and\ \citenamefont {Cabello}}]{Bharti:2022PRXQ}%
  \BibitemOpen
  \bibfield  {author} {\bibinfo {author} {\bibfnamefont {K.}~\bibnamefont
  {Bharti}}, \bibinfo {author} {\bibfnamefont {M.}~\bibnamefont {Ray}},
  \bibinfo {author} {\bibfnamefont {Z.-P.}\ \bibnamefont {Xu}}, \bibinfo
  {author} {\bibfnamefont {M.}~\bibnamefont {Hayashi}}, \bibinfo {author}
  {\bibfnamefont {L.-C.}\ \bibnamefont {Kwek}},\ and\ \bibinfo {author}
  {\bibfnamefont {A.}~\bibnamefont {Cabello}},\ }\bibfield  {title} {\bibinfo
  {title} {{Graph-Theoretic Approach for Self-Testing in Bell Scenarios}},\
  }\href {https://doi.org/10.1103/PRXQuantum.3.030344} {\bibfield  {journal}
  {\bibinfo  {journal} {PRX Quantum}\ }\textbf {\bibinfo {volume} {3}},\
  \bibinfo {pages} {030344} (\bibinfo {year} {2022})}\BibitemShut {NoStop}%
\end{thebibliography}%

\cleardoublepage
\appendix

\section{Optimality of the strategy and nonlocal content of the experimental correlations}


Here, we prove that the adopted strategy is optimal. That is, according to quantum mechanics, no strategy permits one to win the odd-cycle game with a higher probability. We also show how to compute the nonlocal content of our experimental correlations and compare them with the theoretical predictions for an ideal experiment.

Given a nonsignalling correlation for a bipartite Bell scenario, $P(a,b|x,y)$, where $x$ and $y$ are Alice's and Bob's inputs, respectively, and $a$ and $b$ are Alice's and Bob's outputs, respectively, let us consider all possible decompositions
\begin{equation}
    \label{deco}
    P(a,b|x,y) = q_\mathrm{L} p_\mathrm{L}(a,b|x,y) + (1 - q_\mathrm{L})p_\mathrm{NL}(a,b|x,y),
\end{equation}
in terms of arbitrary local and nonlocal (nonsignalling) correlations, $p_\mathrm{L}(a,b|x,y)$ and $p_\mathrm{NL}(a,b|x,y)$, respectively, with respective weights $q_\mathrm{L}$ and $1 - q_\mathrm{L}$, with $0\leq q_\mathrm{L} \leq 1$. The {\em local content} or {\em local fraction} of $P(a,b|x,y)$~\cite{Elitzur:1992PLA} is the maximum local weight over all
possible decompositions of the form (\ref{deco}). That is,
\begin{equation}
    p_\mathrm{L} \doteq \max_{\{p_\mathrm{L},p_\mathrm{NL}\}} q_\mathrm{L}.
\end{equation}
The local content quantifies the fraction of the correlations that can be described by a local model. The {\em nonlocal content} or {\em nonlocal fraction}, defined as
\begin{equation}
    p_\mathrm{NL} \doteq 1-p_\mathrm{L}
\end{equation}
is a measure of the nonlocality of the correlation.

The local content is upper bounded~\cite{Aolita:2012PRA} as follows:
\begin{equation}
    \label{pLbound}
    p_\mathrm{L} \le \frac{\omega_\mathrm{ns} - \omega_\mathrm{q}}{\omega_\mathrm{ns} - \omega_\mathrm{c}},
\end{equation}
where
\begin{equation} \label{Bellineq}
    \sum c_{a,b,x,y} P(a,b|x,y) \le \omega_\mathrm{c}
\end{equation}
is a Bell inequality with corresponding $c_{a,b,x,y} \in \mathbb{R}$, $\omega_\mathrm{c}$ is the classical bound (i.e., the bound for local hidden variable models), and $\omega_\mathrm{ns}$ and $\omega_\mathrm{q}$ are the nonsignalling and quantum bounds for the left-hand side of \eqref{Bellineq}, respectively.
The Bell inequality associated with the winning probability in the odd-cycle game is Eq.~\eqref{winprob}.

\begin{figure}[h!]
    \centering
    \includegraphics[scale = 0.27]{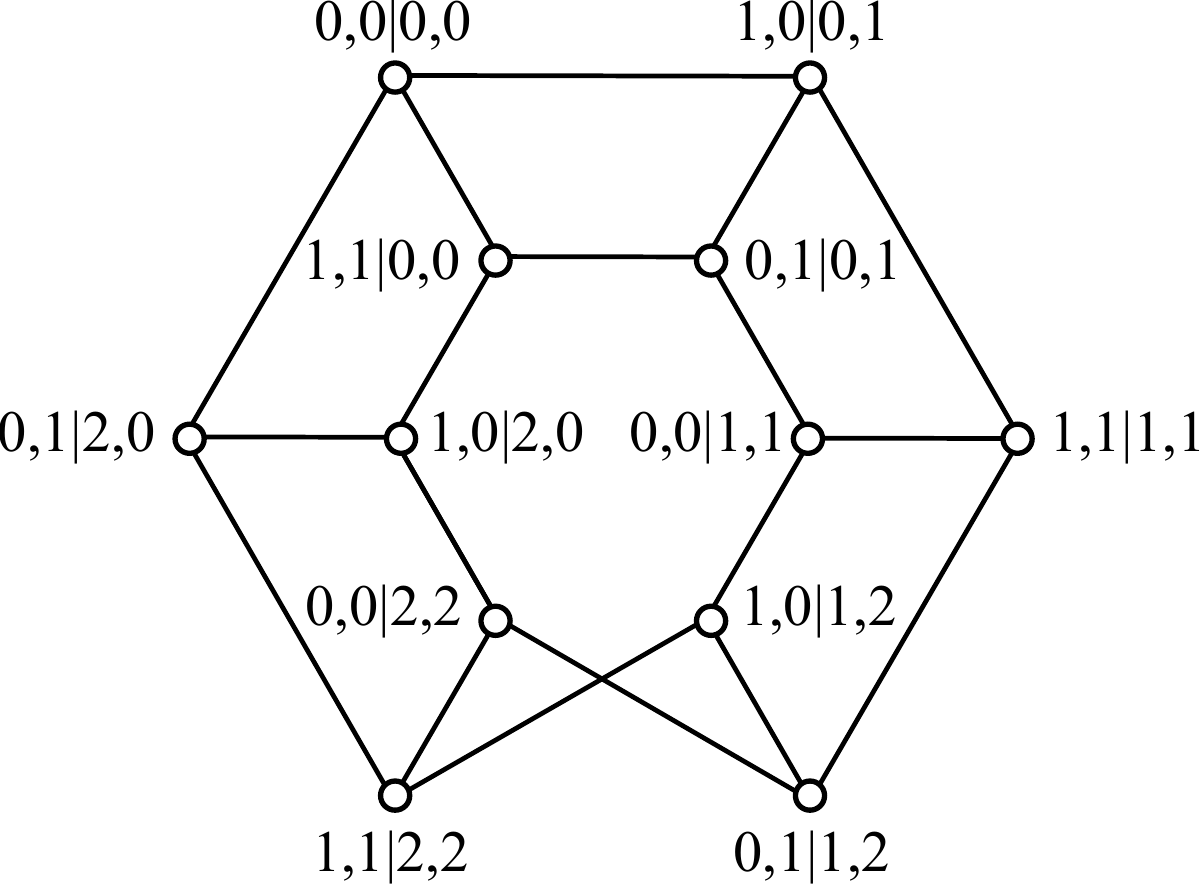}
    \caption{\label{fig5} Graph of exclusivity of the events in the left-hand side of \eqref{winprob} for $n=3$. It is called the M\"obius ladder of order $12$.}
\end{figure}

An elegant way to compute $\omega_\mathrm{c}$, $\omega_\mathrm{q}$, and $\omega_\mathrm{ns}$ is using the graph approach to correlations~\cite{CSWPRL2014} according to which
\begin{subequations}
    \begin{align}
        \omega_\mathrm{c} &= \frac{1}{2 n} \alpha(G_n), \\
        \omega_\mathrm{q} &\le \frac{1}{2 n} \vartheta(G_n), \\
        \omega_\mathrm{ns} &= \frac{1}{2 n} \alpha^\ast (G_n),
    \end{align}
\end{subequations}
where $\alpha(G_n)$, $\vartheta(G_n)$, and $\alpha^\ast(G_n)$ are the independence, Lov\'asz, and fractional packing numbers of $G_n$, respectively~\cite{CSWPRL2014}, and $G_n$ is the graph of exclusivity of the events in the left-hand side of \eqref{winprob}. That is, the graph in which the events are represented by vertices and mutually exclusive events are adjoint. Events $(a,b|x,y)$ and $(a',b'|x',y')$ are mutually exclusive if and only if $x=x'$ and $a \neq a'$ or $y=y'$ and $b \neq b'$. In the case of the left-hand side of \eqref{winprob}, $G_n$ is the M\"obius ladder of order $4 n$, That is,
\begin{equation}
    G_n=M_{4n}.
\end{equation}
For convenience, $G_3=M_{12}$ is shown in Fig.~\ref{fig5}.


In our case,
\begin{subequations}
    \begin{align}
        \alpha(M_{4n})&= 2n-1,\\
        \vartheta(M_{4n})&= n \left[1+ \cos\left(\frac{\pi}{2 n}\right)\right],\\
        \alpha^\ast (M_{4n})&=2n.
    \end{align}
\end{subequations}
The values of $\alpha(M_{4n})$ and $\alpha^\ast (M_{4n})$ are easy to obtain. The value of $\vartheta(M_{4n})$ was conjectured in~\cite{Araujo:2013PRA} [Eq.~(10) and Appendix D] and proven in~\cite{Bharti:2022PRXQ} (Sec.~IV.D). Notice that
\begin{equation}
    n \left[1+ \cos\left(\frac{\pi}{2 n}\right)\right] = 2 n \cos ^2\left(\frac{\pi }{4 n}\right).
\end{equation}
This means that the quantum strategy used in this work is optimal, as it {\em saturates} the quantum bound. This proof of the optimality of the strategy is arguably more elegant than the one in~\cite{cleve_consequences_2004}.

The nonlocal content is therefore
\begin{equation}
    p_\mathrm{NL} \ge 1-\left\{2n-n \left[1+ \cos\left(\frac{\pi}{2 n}\right)\right]\right\},
\end{equation}
which tends to $1$ as $n$ tends to infinity.

\end{document}